\documentclass[preprint,showpacs,showkeys,aps,prd,amsfonts,eqsecnum,nofootinbib,
floatfix]
{revtex4-1}
\usepackage{graphicx,epsfig}
\usepackage[dvips]{color}

\def\gsim{\lower0.5ex\hbox{$\:\buildrel >\over\sim\:$}}
\def\lsim{\lower0.5ex\hbox{$\:\buildrel <\over\sim\:$}}
\begin{document}
%

\title{\Large Higgs Bosons in a minimal R-parity conserving\\ left-right
supersymmetric model}
\author{Mariana Frank and Beste Korutlu }

\affiliation{Department of Physics, Concordia University, 7141 Sherbrooke St. 
West, Montreal, Quebec, CANADA H4B 1R6}
 

\begin{abstract}

We revisit the Higgs sector of the left-right supersymmetric model.  We study
the scalar potential in a version of the model in which the minimum  is the
charge conserving vacuum state,  without $R-$parity violation or additional
non-renormalizable terms in the Lagrangian. We analyze the dependence of the
potential and of the Higgs mass spectrum on the various parameters of the model,
pinpointing the most sensitive ones. We also show that, contrary to previous
expectations, the model can predict light neutral flavor-conserving
Higgs bosons, while the flavor-violating ones are heavy, and within the limits
from $K^0-{\bar K}^0$, $D^0-{\bar D}^0$ and $B_{d,s}^0-{\bar
B}_{d,s}^0$ mixings. We study variants of the model in which at least one pair of
doubly-charged Higgs bosons is light,  and show that the parameter space for
such Higgs masses and mixings is very restrictive, thus making the model more
predictive.

\pacs{12.15.Ji, 12.60.Cn, 12.60.Fr, 12.60.Jv.}
\keywords{Higgs Bosons, Left Right Symmetry, Supersymmetry}
\end{abstract}
\maketitle

\section{Introduction}
\label{sec:intro}
 
Within this decade, the LHC will play a significant role in probing the
Standard Model (SM) of electroweak 
interactions and disentangling the models beyond it. The progress expected in
experimental high energy physics will complement theoretical explorations of
various scenarios of new physics.   The experimental data could confirm 
any of the many theoretical models of new physics advanced  over the last 
decades. 

One of the first observations expected at the LHC is the Higgs boson. This is
the one remaining piece of puzzle missing from the SM and on
this finding rests our understanding of mass generation. However, most models
beyond the SM also predict the existence of one or more Higgs bosons. Some of
them might be heavy, but several are expected to be light. While the standard
model contains one neutral Higgs boson, many models predict one or more Higgs
doublets, thus at least one charged Higgs boson (such as the many variants of the
two Higgs doublet models and supersymmetry). Finding a light charged Higgs boson
would raise problems as to which fundamental gauge symmetry is responsible for
its existence. The hope of a clearer signal rests on more exotic Higgs bosons,
such as the ones predicted in left-right models \cite{Mohapatra:1974gc}.
Left-right symmetric  models with seesaw neutrino mass generation
\cite{Mohapatra:1979ia}
 predict doubly-charged Higgs bosons \cite{Gunion:1989in}, which, if light, 
would give distinctive and spectacular signals at the colliders. 

Including supersymmetry  adds several attractive features to the left-right model
\cite{Francis:1990pi}. Softly-broken supersymmetry resolves some of the
inconsistencies of the standard model: it provides a solution to the gauge
hierachy problem, a natural candidate for weakly-interacting dark matter, and
allows for gauge coupling unification. In addition, the left-right supersymmetric model (LRSUSY) 
accounts for neutrino masses \cite{Mohapatra:1974gc}, parity violation,  offers
a solution to the strong and weak CP violation without introduction of the axion
\cite{Mohapatra:1995xd}, and explains the absence of excessive SUSY CP
violation. Left-right symmetry is favored by many
extra-dimensional models, and many gauge unification scenarios, such as $SO(10)$
\cite{SO10}.

However the model seems to suffer from a serious shortcoming. Minimization of 
the Higgs potential requires either spontaneous $R-$parity breaking by the
vacuum expectation value (VEV) of the right-chiral scalar neutrino
\cite{Kuchimanchi:1993jg}; or introduction of higher scale non-renormalizable
operators \cite{Aulakh:1998nn,Chacko:1997cm}.  Since an attractive
characteristic of the left-right supersymmetric model  is that explicit
$R-$parity breaking is forbidden by the symmetry of the model,  spontaneous
breaking is not a desirable feature. Ditto for higher order operators at the
Planck scale. The shortcoming comes from the fact that, in the simplest
version of the model, the global minimum  of the theory breaks electric charge,
making the theory unacceptable. This can be remedied by allowing a VEV for the
right sneutrino. The Higgs boson spectrum was previously analyzed in this
variant of the model with $R-$parity violation where
sneutrinos and sleptons mix with the Higgs bosons \cite{Huitu:1994zm}. 

However, a new version of the theory suggested by Babu and Mohapatra
\cite{Babu:2008ep}, allows for both $R-$parity conservation {\it and} the
absence of higher-dimensional operators by inclusion of the Yukawa coupling of
the heavy Majorana neutrino in the effective Lagrangian. We study the Higgs
sector of such a model and examine the masses of the doubly-charged, singly
charged and neutral bosons (both scalar and pseudoscalar sectors). Although the
model depends on many parameters, we show that the masses are sensitive to only
a few, and thus  the model is more predictive. Light doubly-charged Higgs bosons
emerge naturally. The LRSUSY model predicts neutral scalar and pseudoscalar
Higgs bosons that violate  flavor at tree level. We impose conditions coming
from phenomenology:  $K^0-{\bar K}^0$, $D^0-{\bar D}^0$ and
$B_{d,s}^0-{\bar B}_{d,s}^0$ mixing. We show that, contrary to previous
expectations, one can have light neutral and charged Higgs bosons that conserve flavor,
while the flavor violating bosons are in the 600 GeV- 100 TeV scale, as required
by meson mixing constraints. We pinpoint the parameters that the masses are most
sensitive to, and show that they satisfy the constraints in a limited range of
these parameters. We set up the structure of the Higgs potential, masses and
mixing, including the constraints, while leaving the study of the characteristic
signals at the LHC  for a future study. 

The paper is organized as follows. In Section {\ref{lrsusy}} we summarize the
particular LRSUSY model we use, with emphasis on the Higgs structure. In the following
Section ({\ref{results}}) we present analytic formulas for the mass matrices in
the neutral, singly-charged and doubly-charged sectors. In Section
{\ref{constraints}} we present the results of the constraints from 
$K^0-{\bar K}^0$, $D^0-{\bar D}^0$ and $B_{d,s}^0-{\bar B}_{d,s}^0$ mixings on
the Higgs masses and mixings. We illustrate our results by showing two numerical
scenarios for desirable Higgs mass values for the model which satisfy the
constraints in Section {\ref{discussion}}  as well as presenting plots
 for masses consistent with the constraints. We summarize our findings
and conclude  in Section {\ref{conclusion}}.

\section{R-parity Conserving Left-Right Supersymmetric Model}
\label{lrsusy}
  
  The supersymmetric left-right model incorporates supersymmetry in the left-right
model based on the gauge symmetry $SU(3)_c \otimes SU(2)_L \otimes SU(2)_R
\otimes U(1)_{B-L}$. Including the ${B-L}$ (where $B$ and $L$ stand for
baryon and lepton numbers) in a gauge symmetry, the only quantum number left
ungauged in SM,  is an additional attractive feature of the model. The model
contains left and right fermion doublets, as well as triplet gauge bosons for
$SU(2)_L$ and $SU(2)_R$, and a neutral gauge boson for $U(1)_{B-L}$. $R$-parity,
defined as $R_P =
(-1)^{3(B-L)+2s}$ (with $s$ the spin of the
particle), is imposed in the Minimal Supersymmetric Standard Model (MSSM) to avoid dangerous baryon and lepton number
violating
operators, otherwise explicit Yukawa terms that violate $R$-parity can exist in the
Lagrangian.   This explicit $R$-parity breaking is forbidden in LRSUSY models by the
symmetries of the model. In early left-right symmetric models
$SU(2)_R$ doublets were used to break the gauge symmetry. Later $SU(2)_{L,R}$
triplets were introduced to provide the seesaw mechanism for neutrino masses \cite{Mohapatra:1979ia},
and both left- and right-handed triplet Higgs bosons are required by parity
conservation. The model was described extensively in several previous works
\cite{Francis:1990pi}. However  $R$-parity may not be conserved in this setup.
The reason is that the minimum of the potential  prefers a solution in which the
right-chiral scalar
neutrino gets a VEV, thus breaking $R$-parity spontaneously. Two 
scenarios have been proposed which remedy this situation.  One is the model of
Babu and Mohapatra \cite{Babu:2008ep} where an extra singlet Higgs boson is
added to the model and one-loop corrections to the potential show that an
$R$-parity conserving minimum can be found.  The second model is that of Aulakh {\it et. al.} 
 \cite{Aulakh:1997fq}, where the addition of two more triplets,
$\Omega (1, 3, 1, 0)$ and $\Omega_c (1, 1, 3, 0)$, with
zero lepton number,  achieves left-right symmetry breaking with conserved
$R$-parity  at
tree-level. In our work, we adopt the former, as it is a minimal model, and
present a short description below. 
    
  The Higgs sector in this minimal left-right supersymmetric model under the
gauge group, together with
the Higgs VEVs, is given in Table I.

The superpotential of this model is given by
\begin{eqnarray}
 W &=& Y_u Q^T \tau_2 \Phi_1 \tau_2 Q^c +  Y_d Q^T \tau_2 \Phi_2 \tau_2
Q^c + Y_\nu L^T \tau_2 \Phi_1 \tau_2 L^c  + Y_\ell L^T \tau_2
\Phi_2 \tau_2
L^c  +{\rm h.c.} \nonumber \\
&+& i  \left(f^* L^T \tau_2 \Delta L + f L^{cT} \tau_2 {\Delta}^c
L^c
\right) \nonumber \\
&+&  S \left[ \lambda\,{\rm Tr} \left( \Delta^\star \bar{ \Delta}^\star + {\Delta}^c
\bar{ \Delta}^c
\right) +\lambda_{ij} {\rm Tr} \left( \Phi_i^T \tau_2 \Phi_j \tau_2
\right)- {\cal M}_R^2 \right] + W'
\end{eqnarray}
where
\begin{equation}
W' = \left[M_\Delta {\rm Tr}(\Delta  \bar{\Delta}) +
M_\Delta^* {\rm Tr}(\Delta^c  \bar{\Delta}^c)\right] + \mu_{ij} {\rm Tr} \left(
\Phi_i^T \tau_2 \Phi_j \tau_2 \right) + {\cal M}_S S^2
+ \lambda_S S^3~.
\end{equation}
Here $Y_{u,d}$ and $Y_{\nu,\ell}$ in Eq. (2.1) are quark and lepton
Yukawa coupling matrices, while $f$ is the Majorana neutrino
Yukawa coupling. We choose to work with $W'=0$, which leads to an enhanced
$R-$symmetry and a natural interpretation of the supersymmetric $\mu$ term, as
explained below. 

\begin{table}
\caption{Higgs sector in the Minimal Supersymmetric Left-Right Model}
\centering
\small{
\begin{tabular}{|c|c|c|}
\hline
 & & \\ 
Higgs Field & Matrix Representation & Vacuum Expectation Values\\
 & & \\ 
\hline
 & & \\ 
$\Delta(1,3,1,2)$ &
$\left(
\begin{array}{cc}
\frac{\delta^{+}}{\sqrt{2}} & \delta^{++}\\
\delta^{0} & -\frac{\delta^{+}}{\sqrt{2}}
\end{array}
\right)$ & 
$\left(
\begin{array}{cc}
0 & 0\\
v_{L} & 0
\end{array}
\right)$\\
 & & \\ 
\hline
 & & \\ 
$\bar{\Delta}(1,3,1,-2)$ &
$\left(
\begin{array}{cc}
\frac{\bar{\delta}^{-}}{\sqrt{2}} & \bar{\delta}^{0}\\
\bar{\delta}^{--} & -\frac{\bar{\delta}^{-}}{\sqrt{2}}
\end{array}
\right)$ & $\left(
\begin{array}{cc}
0 & \bar{v}_{L}\\
0 & 0
\end{array}
\right)$\\
 & & \\ 
\hline
 & & \\ 
$\Delta^{c}(1,1,3,-2)$ &
$\left(
\begin{array}{cc}
\frac{\delta^{c^{-}}}{\sqrt{2}} & \delta^{c^{0}}\\
\delta^{c^{--}} & -\frac{\delta^{c^{-}}}{\sqrt{2}}
\end{array}
\right)$ & $\left(
\begin{array}{cc}
0 & v_{R}\\
0 & 0
\end{array}
\right)$\\
 & & \\ 
\hline
 & & \\ 
$\bar{\Delta}^{c}(1,1,3,2)$ & $\left(
\begin{array}{cc}
\frac{\bar{\delta}^{c^{+}}}{\sqrt{2}} & \bar{\delta}^{c^{++}}\\
\bar{\delta}^{c^{0}} & -\frac{\bar{\delta}^{c^{+}}}{\sqrt{2}}
\end{array}
\right)$ & $\left(
\begin{array}{cc}
0 & 0\\
\bar{v}_{R} & 0
\end{array}
\right)$\\
 & & \\ 
\hline
 & & \\ 
$\Phi_{1}(1,2,2,0)$ &
$\left(
\begin{array}{cc}
\phi^{+}_{1} & \phi^{0}_{2}\\
\phi^{0}_{1} & \phi^{-}_{2}
\end{array}
\right)$ & $
\left(
\begin{array}{cc}
0 & \kappa_{1}^\prime\\
\kappa_{1} & 0
\end{array}
\right)$\\
 & & \\ 
\hline
 & & \\ 
$\Phi_{2}(1,2,2,0)$ & 
$\left(
\begin{array}{cc}
\chi^{+}_{1} & \chi^{0}_{2}\\
\chi^{0}_{1} & \chi^{-}_{2}
\end{array}
\right)$ & $\left(
\begin{array}{cc}
0 & \kappa_{2}\\
\kappa_{2}^\prime & 0
\end{array}
\right)$\\
 & & \\ 
\hline
 & & \\ 
S (1,1,1,0) & $S_0$ & $\langle S \rangle$ \\
 & & \\ 
\hline
\end{tabular}}
\end{table}
\label{table:Higgs}

The model is minimal in the following sense:  $\Delta^{c}$ and
$\bar{\Delta}^{c}$ fields are needed for breaking 
$SU(2)_{R}\otimes U(1)_{B-L}$ symmetry without $R$-parity violation, the 
$\Delta$ and $\bar{\Delta}$ fields are for parity invariance,  and the 
two bidoublets $\Phi_{1}$ and $\Phi_{2}$ are needed to generate the quark and 
lepton masses and Cabibbo Kobayashi Maskawa (CKM) mixings. The singlet field $S$ is introduced to so that 
$SU(2)_{R}\otimes U(1)_{B-L}$ symmetry breaking occurs in the supersymmetric
limit. The charge is defined as 
$$Q=I_{3L}+I_{3R}+\frac{B-L}{2}$$
The VEVs of the Higgs fields in this model needed to break the symmetries as
described above, are
given in Table 1.  If we assume that the VEVs of the bidoublet Higgs are real, the
fermion mass matrices become Hermitian. The VEVs of the left--handed triplet
fields $\Delta, 
\bar{\Delta}$, which determine the tree-level left-handed neutrino masses must be extremely small and are
assumed to be zero.

 In the supersymmetric limit, the
VEV of the singlet $S$ Higgs boson is zero, but after SUSY breaking,
$\left\langle S \right\rangle \sim m_{\rm SUSY}$.  Thus the $\mu$
term for the bidoublet $\Phi$ will arise from the coupling
$\lambda_{ij}$, with a magnitude of order $m_{\rm SUSY}$ \cite{Babu:2008ep}. 
In the SUSY limit,
\begin{equation}
|v_R| = |\overline{v}_R|,~~\lambda v_R \overline{v}_R = {\cal M}_R^2,~~
\left\langle S \right\rangle = 0~.
\end{equation}
The VEV of $S$ field,  generated after SUSY breaking, arises from  linear
terms in SUSY breaking
\begin{equation}
V_{\rm soft} = A_\lambda \lambda S {\rm Tr}(\Delta^c \bar{\Delta}^c
+\Delta^\star \bar{\Delta}^\star
) - C_\lambda {\cal M}_R^2 S + {\rm h.c.}
\end{equation}
Minimization of the resulting potential yields 
$\left\langle S^* \right \rangle = {1 \over 2\lambda} (C_\lambda -
A_\lambda)~$, 
which is of order $m_{\rm SUSY}$.  If the coupling
$\lambda$ is  small, then  $\left\langle S
\right\rangle$ can be above the SUSY breaking scale.  This feature
can be used to make one pair of Higgs doublet superfields 
heavier than the SUSY breaking scale. However, the masses of
doubly charged fermionic fields, which are equal to
$\lambda\left\langle S \right \rangle$ must remain below a TeV. Consistency of
the model  (non-vanishing CKM mixing angle) requires the  asymmetry  $\mu_{12} =
\mu_{21}$. 

The full potential of the model relevant for symmetry breaking includes
$F$-term, $D$-term and soft SUSY breaking contributions.  They are given by
\begin{eqnarray}
V_{F}&=&\left|\lambda{\rm Tr}(\Delta^{*}\bar{\Delta}^{*}+
\Delta^{c}\bar{\Delta}^{c})+\lambda_{ij}{\rm Tr}(\Phi_{i}^{T}\tau_{2}\Phi_{j}
\tau_{2})-{\cal M}_{R}^{2}\right|^2\nonumber\\
&+& \lambda^{2}|S|^{2}\left|{\rm Tr}(\Delta^{*}\Delta^{*\dagger})+{\rm
Tr}(\bar{\Delta}^{*}\bar{\Delta}^{*\dagger})+{\rm
Tr}(\Delta^{c}\Delta^{c\dagger})+{\rm
Tr}(\bar{\Delta}^{c}\bar{\Delta}^{c\dagger})\right|,
\nonumber\\
V_{\rm soft} &=& M_{1}^{2}{\rm Tr}
(\Delta^{*\dagger}\Delta^{*}+\Delta^{c\dagger}\Delta^{c})+
M_{2}^{2}{\rm
Tr}(\bar{\Delta}^{*\dagger}\bar{\Delta}^{*}+\bar{\Delta}^{c\dagger}\bar{\Delta}^
{c})\nonumber\\
&+&M_{3}^{2}\Phi_{1}^{\dagger}\Phi_{1}+M_{4}^{2}\Phi_{2}^{\dagger}
\Phi_{2}+M_{S}^{2}|S|^{2}\nonumber\\
&+& \{A_{\lambda}\lambda
S{\rm Tr}(\Delta^{*}\bar{\Delta}^{*}+\Delta^{c}\bar{\Delta}^{c})-
C_{\lambda}{\cal M}_R^2 S + h.c.\}, \nonumber\\
V_D &=&\frac{g_{L}^{2}}{8}\sum_{i}\left|{\rm Tr}(2
\Delta^{*\dagger}\tau_{i}\Delta^{*} +
2\bar{\Delta}^{*\dagger}\tau_{i}\bar{\Delta}^{*}+\Phi_{a}\tau_{i}^{T}
\Phi_{b}^{\dagger})\right|^2\nonumber\\
&+&\frac{g_{R}^{2}}{8}\sum_{i}\left|{\rm Tr}(2\Delta^{c\dagger}\tau_{i}
\Delta^{c}+2\bar{\Delta}^{c\dagger}\tau_{i}\bar{\Delta}^{c}+\Phi_{a}\tau_{i}^{T}
\Phi_{b}^{\dagger})\right|^2\nonumber\\
&+& {\frac{g'^{2}}{2}} \left|{\rm
Tr}(-\Delta^{*\dagger}\Delta^{*}+\bar{\Delta}^{*\dagger}\bar{\Delta}^{*}-\Delta^
{c\dagger}\Delta^{c}+\bar{\Delta}^{c\dagger}\bar{\Delta}^{c})\right|^2.
\end{eqnarray}
All terms in the scalar potential are identical for the configurations in which
VEVs are given to the neutral right-handed triplet Higgs, or the charged Higgs,
except for the $D-$term, which is lower for the charge breaking configuration. 
Previous solutions suggested are breaking $R-$parity, which would have the
attractive feature that $v_R \sim 1$ TeV, but which abandons the LSP as the
candidate for dark matter \cite{Kuchimanchi:1993jg}; or introducing higher
dimensional operators to lower the charge conserving vacuum, with $v_R \sim
10^{11}$ GeV, but loosing the solution to strong and weak CP violation
\cite{Aulakh:1997fq}. More recently, a new version of the model
\cite{Babu:2008ep} examined effective operators which generate terms of the form
${\rm Tr}(\Delta^c \Delta^c) {\rm Tr} (\Delta^{c \dagger} \Delta^{c \dagger})$ at one loop
level, induced by the Majorana neutrino $\nu^c$ couplings with $\Delta^c$. These
operators mimic the effects of the higher dimensional operators in previous
versions, without the need to introduce them explicitly, thus solving the
problem of the global minimum (if their coefficient is positive). The advantage
of such a formalism is that the masses are very predictive, as they do not
depend on coefficients of ad-hoc higher order terms, or sneutrino VEVs. In the next section, we
study explicitly the implications for the Higgs masses in this model. 


\section{Higgs Boson Composition and Masses}
\label{results}

The Higgs boson spectrum was previously analyzed in a variant of the model
\cite{Huitu:1994zm} with $R-$parity violation. The new features of the present analysis
are 1) we employ a version of the model that uses the right-chiral neutrino
couplings to the triplet Higgs bosons to eliminate the need for $L$-number
violation; and 2) we include constraints from FCNC processes to predict the
range of Higgs masses and parameters in LRSUSY. Effectively, we are looking at a very
different model and Higgs sector than in \cite{Huitu:1994zm}.

We proceed the usual way to find the masses and mixing matrices for the
Higgs bosons in this model. We minimize the Higgs potential given in the previous
section, taking into account corrections induced by the heavy Majorana neutrino
Yukawa couplings. This insures that the minimum of the potential is charge
conserving. We forgo providing explicit expressions for the equations obtained
by taking 
$$\frac {\partial V}{\partial \kappa_1}=  \frac{\partial V}{\partial
\kappa_2}=\frac{\partial V}{\partial v_L}=\frac{\partial V}{\partial {\bar
v}_L}=\frac{\partial V}{\partial v_R}=\frac{\partial V}{\partial {\bar
v}_R}=\frac {\partial V}{\partial \langle S \rangle }=0,$$
instead we give the relevant mass matrices for the Higgs fields. For simplicity,
we use the abbreviations  
\begin{eqnarray}
\kappa^{2}_{dif}&=&\kappa^{2}_{1}-\kappa^{2}_{2},\\
\rho^{2}_{dif}&=&v^{2}_{R}-\bar{v}^{2}_{R}+\frac{1}{2×}(\kappa^{2}_{1}-\kappa^{
2}_{2}),\\
Y&=&A_{\lambda}\lambda
S+\lambda(-M^{2}_{R}-2\lambda_{21}\kappa_{1}\kappa_{2}+\lambda
v_{R}\bar{v}_{R}),\\
M&=&2\lambda_{21}(-M^{2}_{R}-2\lambda_{21}\kappa_{1}\kappa_{2}
+\lambda v_{R}\bar{v}_{R}),\\
f(\epsilon)&=&\epsilon(\frac{M}{2\lambda_{21}}-2\lambda_{21}\kappa_{1}
\kappa_{2}-\epsilon\kappa_{1}\kappa_{2}),\\
g(\epsilon)&=&\epsilon\lambda\kappa_{1}\kappa_{2},\\
h(\epsilon)&=&\epsilon\kappa_{1}\kappa_{2}(4\lambda_{21}+\epsilon),
\end{eqnarray}
with $\epsilon=\mu_{21}-\mu_{12}$,  small but non-zero after symmetry breaking.
\subsection{Doubly Charged Higgs Boson Masses}

Mass matrices for the doubly charged Higgs fields are of block diagonal form of
one two by two matrix for $(\delta^{^{++}},\bar{\delta}^{^{--*}})$ fields and one two
by two matrix for $(\delta^{c^{--*}},\bar{\delta}^{c^{++}})$ fields respectively,
\begin{eqnarray}
M^{2}_{\delta^{++}\bar{\delta}^{--*}}&=&
\left(
\begin{array}{cc}
\frac{1}{2}g^{2}_{L}\kappa^{2}_{dif}-g^{2}_{R}\rho^
{2}_{dif}
-\frac{\bar{v}_{R}}{v_{R}}Y' &
Y'\\
Y' &
-\frac{1}{2}g^{2}_{L}\kappa^{2}_{dif}+g^{2}_{R}\rho^
{2}_{dif}
-\frac{v_{R}}{\bar{v}_{R}}Y'
\end{array}
\right),\nonumber\\
M^{2}_{\delta^{c^{--*}}\bar{\delta}^{c^{++}}}&=&
\left(
\begin{array}{cc}
-2g^{2}_{R}\rho^{2}_{dif}-\frac{\bar{v}_{R}}{v_{R}}Y' &
Y'\\
Y' &
2g^{2}_{R}\rho^{2}_{dif}-\frac{v_{R}}{
\bar{v}_{R}}Y'
\end{array}
\right),
\end{eqnarray}
where $Y'=Y-g(\epsilon)$. 
From these expressions we can find the exact analytic forms for the doubly charged Higgs
masses. In the limit $v_R \simeq \bar{v}_R$, these are:
\begin{eqnarray}
M^2_{H^{++}_{L\,1,2}}&\simeq &- Y^\prime  \pm \frac12 \sqrt{
\left(\frac{1}{2}g^{2}_{L}\kappa^{2}_{dif}-g^{2}_{R}\rho^
{2}_{dif}\right )^2+2Y^{\prime 2}} \nonumber\\
M^2_{H^{++}_{R\,1,2}}&\simeq&-Y^\prime \pm \frac12 \sqrt{ 4 g^{4}_{R}\rho^
{4}_{dif}+2Y^{\prime 2}} 
\end{eqnarray}
Thus, in all cases, the left-handed doubly charged Higgs fields are expected to
be lighter than the right-handed ones.
\subsection{Singly Charged Higgs Boson Masses}

Mass matrices for the singly charged Higgs fields are
of block diagonal form of
one two by two matrix for $(\delta^{+},\bar{\delta}^{-*})$ fields,
one two by two matrix for $(\phi^{+}_{1},\chi^{-*}_{2})$ fields and one four by
four matrix 
for
$(\delta^{c^{+}},\bar{\delta}^{c^{-*}},\phi^{-*}_{2},\chi^{+}_{1})$ fields
respectively,
\begin{equation}
M^{2}_{\delta^{+}\bar{\delta}^{-*}}=
\left(
\begin{array}{cc}
-g^{2}_{R}\rho^{2}_{dif}-\frac{\bar{v_{R}}}{{v}_{R}}Y' & Y'\\
Y' & g^{2}_{R}\rho^{2}_{dif}-\frac{v_{R}}{\bar{v}_{R}}Y'
\end{array}
\right),
\end{equation}

\begin{equation}
M^{2}_{\phi^{+}_{1},\chi^{-*}_{2}}=
\left(
\begin{array}{cc}
\frac{\kappa_{2}}{\kappa_{1}}M' & M' \\
M' & \frac{\kappa_{1}}{\kappa_{2}}M'
\end{array}
\right),
\end{equation}
where $M'=M+f(\epsilon)$. The elements of the four by four matrix are
\begin{eqnarray}
M^{2}_{\delta^{c^{-*}}\delta^{c^{-}}}&=&g^{2}_{R}v^{2}_{R}-g^{2}_{R}
\rho^{2}_{dif}-\frac{\bar{v}_{R}}{v_{R}}Y'
\\
M^{2}_{\delta^{c^{-*}}\bar{\delta}^{c^{+*}}}&=&-g^{2}_{R}v_{R}\bar{v}_{R}+Y'\\
M^{2}_{\delta^{c^{-*}}\phi^{-}_{2}}&=&-\frac{1}{\sqrt{2}}g^{2}_{R}\kappa_{1}v_{
R
}\\
M^{2}_{\delta^{c^{-*}}\chi^{+*}_{1}}&=&-\frac{1}{\sqrt{2}}g^{2}_{R}\kappa_
{2}v_{R}\\
M^{2}_{\bar{\delta}^{c^{+}}\bar{\delta}^{c^{+*}}}&=&g^{2}_{R}\bar{v}^{2}_{R}+g^{
2 } _ { R}\rho^{2}_{dif}-\frac{v_{R}}{\bar{v}_
{R}}Y'\\
M^{2}_{\bar{\delta}^{c^{+}}\phi^{-}_{2}}&=&\frac{1}{\sqrt{2}}g^{2}_{R}
\kappa_{1}\bar{v}_ { R
}\\
M^{2}_{\bar{\delta}^{c^{+}}\chi^{+*}_{1}}&=&\frac{1}{\sqrt{2}}g^{2}_{R}
\kappa_{2}\bar{v}_{ R
}\\
M^{2}_{\phi^{-*}_{2}\phi^{-}_{2}}&=&\frac{1}{2}\kappa^{2}_{1}(g^{2}_{L}
+g^{2}_{R})-\frac{1}{2}g^{2}_{L}\kappa^{2}_{dif}-g^{2}_{R}\rho^{2}_{dif}
+\frac{\kappa_{2}}{\kappa_{1}}M'\\
M^{2}_{\phi^{-*}_{2}\chi^{+*}_{1}}&=&\frac{1}{2}\kappa_{1}\kappa_{2}(g^{2}_{L}
+g^ {
2}_{R})+M'\\
M^{2}_{\chi^{+}_{1}\chi^{-}_{1}}&=&\frac{1}{2}\kappa^{2}_{2}(g^{2}_{L}
+g^{2}_{R})+\frac{1}{2}g^{2}_{L}\kappa^{2}_{dif}+g^{2}_{R}\rho^{2}_{dif}
+\frac{\kappa_{1}}{\kappa_{2}}M'
\end{eqnarray}

\subsection{Neutral Higgs Boson Masses}

Mass matrices for the neutral scalar Higgs fields are of block diagonal form of
one two by two matrix for $(\delta^{0r},\bar{\delta}^{0r})$ fields,
one two by two matrix for $(\phi^{0r}_{2},\chi^{0r}_{1})$ fields and one five by
five
matrix for 
$(\delta^{c^{0r}},\bar{\delta}^{c^{0r}},\phi^{0r}_{1},\chi^{0r}_{2}, S^{0r})$
fields
respectively,
\begin{eqnarray}
M^{2}_{\delta^{0r}\bar{\delta}^{0r}}&=& \left(
\begin{array}{cc}
-\frac{1}{2}g^{2}_{L}\kappa^{2}_{dif}-g^{2}_{R}\rho^{2}_{dif}
-\frac{\bar{v}_{R}}{v_{R}}Y' & Y'\\
Y'& \frac{1}{2}g^{2}_{L}\kappa^{2}_{dif}+g^{2}_{R}\rho^{2}_{dif}
-\frac{v_{R}}{\bar{v}_{R}}Y'
\end{array}
\right)\nonumber\\
 M^{2}_{\phi^{0r}_{2},\chi^{0r}_{1}}&=& \left(
\begin{array}{cc}
-\frac{1}{2}g^{2}_{L}\kappa^{2}_{dif}-g^{2}_{R}\rho^{2}_{
dif}+\frac{\kappa_{2}}{\kappa_{1}}M' & -M' \\
-M'& \frac{1}{2}g^{2}_{L}\kappa^{2}_{dif}+g^{2}_{R}\rho^{2}_{
dif}+\frac{\kappa_{1}}{\kappa_{2}}M'
\end{array}
\right).
\end{eqnarray}
The elements of the five by five matrix are
\begin{eqnarray}
M^{2}_{\delta^{c^{0r}}\delta^{c^{0r}}}&=&2v^{2}_{R}(g^{2}_{
B-L}+g^{2}_{R})+\lambda^{2}\bar{v}^{2}_{R}-\frac{\bar{v}_{R}}{v_{R}}Y'\\
M^{2}_{\delta^{c^{0r}}\bar{\delta}^{c^{0r}}}&=&-2v_{R}
\bar{v}_{R}(g^{2}_{B-L}+g^{2}_{R})+\lambda^{2}v_{R}\bar{v}_{R}+Y'\\
M^{2}_{\delta^{c^{0r}}\phi^{0r}_{1}}&=&g^{2}_{R}\kappa_{1}v_{R}
-2\lambda\lambda_{21}\kappa_{2}\bar{v}_{R}-2\frac{\bar{v}_{R}}{\kappa_{1}}
g(\epsilon)\\
M^{2}_{\delta^{c^{0r}}\chi^{0r}_{2}}&=&-g^{2}_{R}\kappa_{2}v_{R}
-2\lambda\lambda_{21}\kappa_{1}\bar{v}_{R}-\frac{\bar{v}_{R}}{\kappa_{2}}
g(\epsilon)\\
M^{2}_{\delta^{c^{0r}}S^{0r}}&=&2\lambda^{2}S
v_{R}+A_{\lambda}\lambda\bar{v}_{R}\\
M^{2}_{\bar{\delta}^{c^{0r}}\bar{\delta}^{c^{0r}}}&=&2(g^{2}_{B-L}+g^{2}_{R}
)\bar {v}^{2}_{R}
+\lambda^{2}v^{2}_{R}-\frac{v_{R}}{\bar{v}_{R}}Y'\\
M^{2}_{\bar{\delta}^{c^{0r}}\phi^{0r}_{1}}&=&-g^{2}_{R}\kappa_{1}\bar{v}_{R}
-2\lambda\lambda_{21}\kappa_{2}v_{R}-\frac{v_{R}}{\kappa_{1}}g(\epsilon)\\
M^{2}_{\bar{\delta}^{c^{0r}}\chi^{0r}_{2}}&=&g^{2}_{R}\kappa_{2}\bar{v}_{R}
-2\lambda\lambda_{21}\kappa_{1}v_{R}-\frac{v_{R}}{\kappa_{2}}g(\epsilon)\\
M^{2}_{\bar{\delta}^{c^{0r}}S^{0r}}&=&2\lambda^{2}S\bar{v}_{R}+A_{\lambda}
\lambda v_{R}\\
M^{2}_{\phi^{0r}_{1}\phi^{0r}_{1}}&=&\frac{1}{2}\kappa^{2}_{1}(g^{2}_{L}
+g^{2}_{R})+4\lambda^{2}_{21}\kappa^{2}_{2}
+\frac{\kappa_{2}}{\kappa_{1}}[M'+h(\epsilon)]\\
M^{2}_{\phi^{0r}_{1}\chi^{0r}_{2}}&=&-\frac{1}{2}\kappa_{1}\kappa_{
2}(g^{2}_{L}+g^{2}_{R})+4\lambda^{2}_{21}\kappa_{1}\kappa_{2}
-[M'-h(\epsilon)]\\
M^{2}_{\phi^{0r}_{1}S^{0r}}&=&0\\
M^{2}_{\chi^{0r}_{2}\chi^{0r}_{2}}&=&\frac{1}{2}\kappa^{2}_{2}(g^{2}_{L}
+g^{2}_{R})+4\lambda^{2}_{21}\kappa^{2}_{1}
+\frac{\kappa_{1}}{\kappa_{2}}[M'+h(\epsilon)]\\
M^{2}_{\chi^{0r}_{2}S^{0r}}&=&0\\
M^{2}_{S^{0r}S^{0r}}&=&M_{S}^{2}+\lambda^{2}(v_{R}^{2}+\bar{v}_{R}^{2})
\end{eqnarray}
%
Mass matrices for the neutral pseudoscalar Higgs fields are similar of block
diagonal form of one two by two matrix for $(\delta^{0i},\bar{\delta}^{0i})$
fields, one two by two matrix for $(\phi^{0i}_{2},\chi^{0i}_{1})$ fields and
one five by five matrix for
$(\delta^{c^{0i}},\bar{\delta}^{c^{0i}},\phi^{0i}_{1},\chi^{0i}_{2},
S^{0i})$ fields respectively,
\begin{eqnarray}
M^{2}_{\delta^{0i}\bar{\delta}^{0i}}&=& \left(
\begin{array}{cc}
-\frac{1}{2}g^{2}_{L}\kappa^{2}_{dif}-g^{2}_{R}\rho^{2}_{dif}
-\frac{\bar{v}_{R}}{v_{R}}Y' & -Y'\\
-Y' & \frac{1}{2}g^{2}_{L}\kappa^{2}_{dif}+g^{2}_{R}\rho^{2}_{dif}
-\frac{v_{R}}{\bar{v}_{R}}Y'
\end{array}\right),\nonumber\\
M^{2}_{\phi^{0i}_{2},\chi^{0i}_{1}}&=& \left(
\begin{array}{cc}
-\frac{1}{2}g^{2}_{L}\kappa^{2}_{dif}-g^{2}_{R}\rho^{2}_{dif}+\frac{\kappa_{2}
}{\kappa_{1}}M' & M' \\
M' &
\frac{1}{2}g^{2}_{L}\kappa^{2}_{dif}+g^{2}_{R}\rho^{2}_{dif}+\frac{\kappa_{1}}
{\kappa_{2}}M'
\end{array}
\right)
\end{eqnarray}
The elements of the five by five matrix are
\begin{eqnarray}
M^{2}_{\delta^{c^{0i}}\delta^{c^{0i}}}&=&\lambda^{
2}\bar{v}^{2}_{R}-\frac{\bar{v}_{R}}{v_{R}}Y'\\
M^{2}_{\delta^{c^{0i}}\bar{\delta}^{c^{0i}}}&=&\lambda^{2}v_{R}\bar{v}_{R}
-Y'\\
M^{2}_{\delta^{c^{0i}}\phi^{0i}_{1}}&=&-2\lambda\lambda_{21}\kappa_{2}\bar{v}_{R
}
-\frac{\bar{v}_{R}}{\kappa_{1}}g(\epsilon)\\
M^{2}_{\delta^{c^{0i}}\chi^{0i}_{2}}&=&-2\lambda\lambda_{21}\kappa_{1}\bar{v}_{R
}-\frac{\bar{v}_{R}}{\kappa_{2}}g(\epsilon)\\
M^{2}_{\delta^{c^{0i}}S^{0i}}&=&-A_{\lambda}\lambda\bar{v}_{R}\\
M^{2}_{\bar{\delta}^{c^{0i}}\bar{\delta}^{c^{0i}}}&=&\lambda^{2}v^{2}_{R}-
\frac{v_{R}}{\bar{v}_{R}}Y'\\
M^{2}_{\bar{\delta}^{c^{0i}}\phi^{0i}_{1}}&=&-2\lambda\lambda_{21}\kappa_{2}v_{R
}-\frac{v_{R}}{\kappa_{1}}g(\epsilon)\\
M^{2}_{\bar{\delta}^{c^{0i}}\chi^{0i}_{2}}&=&-2\lambda\lambda_{21}\kappa_{1}v_{R
}-\frac{v_{R}}{\kappa_{2}}g(\epsilon)\\
M^{2}_{\bar{\delta}^{c^{0i}}S^{0i}}&=&-A_{\lambda}\lambda v_{R}\\
M^{2}_{\phi^{0i}_{1}\phi^{0i}_{1}}&=&4\lambda_{21}^{2}\kappa_{2}^{2}+
\frac{\kappa_{2}}{\kappa_{1}}[M'+h(\epsilon)]\\
M^{2}_{\phi^{0i}_{1}\chi^{0i}_{2}}&=&4\lambda_{21}^{2}\kappa_{2}
\kappa_{2}+[M'+h(\epsilon)]\\
M^{2}_{\phi^{0i}_{1}S^{i}}&=&0\\
M^{2}_{\chi^{0i}_{2}\chi^{0r}_{2}}&=&4\lambda_{21}^{2}\kappa_{1}^{2}+
\frac{\kappa_{1}}{\kappa_{2}}[M'+h(\epsilon)]\\
M^{2}_{\chi^{0i}_{2}S^{0i}}&=&0\\
M^{2}_{S^{0i}S^{0i}}&=&M_{S}^{2}+\lambda^{2}(v_{R}^{2}+\bar{v}_{R}^{2})
\end{eqnarray}
%

\section{Constraints on the Higgs sector}
\label{constraints}
 \subsection{Flavor Changing Neutral Higgs Bosons}
 As any model with more than one Higgs doublet, the LRSUSY is plagued by
tree-level FCNC-inducing Higgs bosons \cite{Pospelov:1996fq}. We proceed first
by isolating the flavor-violating and flavor-conserving field combinations, then
proceed to subject them to constraints coming from mixings in the kaon, $B$ and
$D$ neutral meson states. We show more explicitly the expressions for the
down-quark sector; the up-quark sector can be obtained simply by the same
method.
 The Yukawa Lagrangian in the quark sector is given by
\begin{equation}\label{quarkyukawa}
\mathcal{L}_{Y}=\bar{d}_{L}Y_{u}\phi_{2}^{0}d_{R}+\bar{d}_{L}Y_{d}\chi_{2}^{0}d_
{R}
+\bar{u}_{L}Y_{u}\phi_{1}^{0}u_{R}+\bar{u}_{L}Y_{d}\chi_{1}^{0}u_{R}+\textrm{
h.c.} ,
\end{equation}
where $Y_{u}$ and $Y_{d}$ are $3\times 3$ Hermitian matrices in flavor
space. When the bi-doublets acquire the VEV as in Table I, 
 with $\kappa_{1}$, $\kappa_{2}$, $\kappa'_{1}$ and $\kappa'_{2}$ real,  the up and the down type quark mass matrices 
 are given by:
\begin{eqnarray}
M_{u}&=&Y_{u}\kappa_{1}+Y_{d}\kappa'_{2}\nonumber\\
M_{d}&=&Y_{u}\kappa'_{1}+Y_{d}\kappa_{2}.
\end{eqnarray}
Inserting the expressions obtained for $Y_{u}$ and $Y_{d}$ in terms of masses, 
the Yukawa Lagrangian in the down type quark sector reads
\begin{eqnarray}\label{YukawaDownN}
\mathcal{L}^{N}_{Y}(d)&=&\frac{\big[d_{L}^{i*}M_{u}^{ij}d_{R}^{j}(\kappa_{2
}\phi^{0}_{2}-\kappa'_{2}\chi^{0}_{2})
+d_{L}^{i*}M_{d}^{ij}d_{R}^{j}(\kappa_{1}\chi^{0}_{2}-\kappa'_{1}\phi^{0}_{
2})\big]}{\kappa_{1}\kappa_{2} -\kappa'_{1}\kappa'_{2}}\nonumber\\
&+&\frac{\big[d_{R}^{j*}M_{u}^{ij*}d_{L}^{i}(\kappa_{2}\phi^{0*}_{2}
-\kappa'_{2}\chi^{0*}_{2})+d_{R}^{j*}M_{d}^{ij*}d_{L
}^{i}(\kappa_{1}\chi^{0*}_{2}-\kappa'_{1}
\phi^{0*}_{2})\big]}{\kappa_{1}\kappa_{2}-\kappa'_{1}\kappa'_{2}}.
\end{eqnarray}
To obtain the physical states we diagonalize the mass matrices by
the unitary transformations
\begin{eqnarray}
M_{u}^{ij}&=&U_{u}^{ik}\hat{M}^{km}_{u}W^{jm*}_{u}\delta^{km},\nonumber\\
M_{d}^{ij}&=&U_{d}^{ik}\hat{M}^{km}_{d}W^{jm*}_{d}\delta^{km},
\end{eqnarray}
where $\hat{M}_{u}$ and $\hat{M}_{d}$ are diagonal up and down type quark mass
matrices. Since $d_{L}$ and $d_{R}$ are weak eigenstates,  
unitary transformations  convert them into mass eigenstates 
\begin{eqnarray}
d_{L}^{i}&\rightarrow&U^{ij}_{d}d^{j}_{L},\nonumber\\
d_{R}^{i}&\rightarrow&W_{d}^{ij}d^{j}_{R}.
\end{eqnarray}
 We define $U_{d}^{ji*}U_{u}^{ik}=V_{L}^{jk}$ and
$W^{lj*}_{u}W_{d}^{jm}=V_{R}^{lm}$ where $V_{L}$ and $V_{R}$ are the components
of the left-handed and right-handed CKM matrices. Then the Yukawa Lagrangian for down type quark fields is given by
\begin{eqnarray}
\mathcal{L}^{N}_{Y}(d)&=&\frac{d_{L}^{n*}V_{L}^{kn*}\hat{M}^{km}_{u}
V^{ml}_{R}d_{R}^{l}\delta^{km}
(\kappa_{2}\phi^{0}_{2}-\kappa'_{2}\chi^{0}_{2})}{\kappa_{1}\kappa_{2}-\kappa'_{
1}\kappa'_{2}}
+\frac{d_{L}^{n*}\delta^{nk}\hat{M}^{km}_{d}\delta^{ml}d_{R}^{l}\delta^{
km}(\kappa_{
1}\chi^{0}_{2}-\kappa'_{1}\phi^{0}_{2})}{\kappa_{1}\kappa_{2} -\kappa'_
{1}\kappa'_{2}}\nonumber\\
&+&\frac{d_{R}^{n*}V_{R}^{mn*}\hat{M}^{km*}_{u}
V^{kl}_{L}d_{L}^{l}\delta^{
km}(\kappa_{2}\phi^{0*}_{2}-\kappa'_{2}\chi^{0*}_{2})}{\kappa_{1
}\kappa_{2}-\kappa'_{1}\kappa'_{2}}
+\frac{d_{R}^{n*}\delta^{nm}\hat{M}^{km*}_{d}\delta^{kl}d_{L}^{l}\delta^{
km}(\kappa_{
1}\chi^{0*}_{2}-\kappa'_{1}\phi^{0*}_{2})}{\kappa_{1}\kappa_{2} -\kappa'_
{1}\kappa'_{2}},\nonumber\\
\end{eqnarray}
where the up and down mass matrices are Hermitian since the VEVs of
bi-doublets are taken to be real. For simplicity, we assume $V_{L}=V_{R}=V$.
The fields $\phi^{0}_{2}$ and $\chi^{0}_{2}$ are complex. Thus we can isolate
two terms in the Lagrangian,  one flavor violating, and one FCNC-conserving.
Writing the neutral and imaginary parts separately, the FCNC Lagrangian reads
\begin{eqnarray}\label{NReal}
\mathcal{L}_{FCNC}(d)&=&\frac{d_{L}^{n*}V^{kn*}\hat{M}^{kk}_{u}
V^{kl}d_{R}^{l}(\kappa_{2}\phi^{0r}_{2}-\kappa'_{2}\chi^{0r}_{2})}{
\kappa_{1}\kappa_{2}
-\kappa'_{1}\kappa'_{2}}
+\frac{id_{L}^{n*}V^{kn*}\hat{M}^{kk}_{u}
V^{kl}d_{R}^{l}
(\kappa_{2}\phi^{0i}_{2}-\kappa'_{2}\chi^{0i}_{2})}{\kappa_{1}\kappa_{2}
-\kappa'_{1}\kappa'_{2}}\nonumber\\
&+&\frac{d_{R}^{n*}V^{kn*}\hat{M}^{kk*}_{u}V^{kl}d_{R}^{l}
(\kappa_{2}\phi^{0r}_{2}-\kappa'_{2}\chi^{0r}_{2})}{\kappa_{1}\kappa_{2}
-\kappa'_{1}\kappa'_{2}}
-\frac{id_{R}^{n*}V^{kn*}\hat{M}^{kk*}_{u}V^{kl}d_{R}^{l}
(\kappa_{2}\phi^{0i}_{2}-\kappa'_{2}\chi^{0i}_{2})}{\kappa_{1}\kappa_{2}
-\kappa'_{1}\kappa'_{2}},\nonumber\\
\end{eqnarray}
where $\phi^{0r}_{2}$ and $\chi^{0r}_{2}$ are the two of the nine bare scalar
fields and $\phi^{0i}_{2}$ and $\chi^{0i}_{2}$
are the two of the nine bare pseudo-scalar fields appearing
in LRSUSY Lagrangian. The $d-s$ coupling in Eq. (\ref{NReal}) allows
 a $\Delta S=2$ transition at tree level.
To evaluate explicitly, we use the Wolfenstein parametrization
with every parameter expanded as a power series in the
parameter $\lambda=|V_{us}|=0.2246\pm 0.0012$ \cite{Wolfenstein:1983yz}.
\begin{equation}
V=\left(
\begin{array}{ccc}
1-\frac{\lambda^{2}}{2} & \lambda & A\lambda^{3}(\rho-i\eta)\\
-\lambda & 1-\frac{\lambda^{2}}{2} & A\lambda^{2}\\
A\lambda^{3}(1-\rho-i\eta) & -A\lambda^{2} & 1
\end{array}
\right)+\mathcal{O}(\lambda^{4}).
\end{equation}
For $\lambda=0.2246$, $A=0.832$, $\rho=0.130$, and $\eta=0.350$
\cite{Bona,Bona2}
\begin{equation}
V^{kd*}\hat{M}^{kk}_{u}V^{ks}=(m_{u}-m_{c})(\lambda-\frac{\lambda^{3}}{2})-m_{t}
A^{2}\lambda^{5}(1-\rho+i\eta).
\end{equation}
We express the bare scalar $\psi^{0r^{T}}=\left(
\delta^{0r} ~ \bar{\delta}^{0r} ~ \delta^{c0r} ~ \bar{\delta}^{c0r}~ 
\phi^{0r}_{1}  ~ \phi^{0r}_{2} ~ \chi^{0r}_{1} ~ \chi^{0r}_{2} ~ S^{0r}
\right)$ and pseudoscalar Higgs fields  $\psi^{0i^{T}}=\left(
\delta^{0i} ~ \bar{\delta}^{0i} ~ \delta^{c0i} ~ \bar{\delta}^{c0i}~ 
\phi^{0i}_{1} ~ \phi^{0i}_{2} ~ \chi^{0i}_{1} ~ \chi^{0i}_{2} ~ S^{0i}
\right)$ 
as  physical CP even Higgs fields
$H^{0r^{T}}=\left(
H^{0r}_{1} ~ H^{0r}_{2} ~ H^{0r}_{3} ~ H^{0r}_{4} ~ H^{0r}_{5} ~ H^{0r}_{6} ~
H^{0r}_{7} ~ H^{0r}_{8} ~ H^{0r}_{9}
\right)$ and physical CP odd Higgs fields $
H^{0i^{T}}=\left(
H^{0i}_{1} ~ H^{0i}_{2} ~ H^{0i}_{3} ~ H^{0i}_{4} ~ H^{0i}_{5} ~ H^{0i}_{6} ~
H^{0i}_{7} ~ H^{0i}_{8} ~ H^{0i}_{9}
\right)$.
Call $A_{ij}$ the transformation matrix which transforms the bare scalar fields
into the physical CP even ones, and $B_{ij}$ matrix which transforms the bare
pseudo-scalar
fields into the physical CP odd ones: $
H^{0r}_{i}=A_{ij}\psi^{0r}_{j}$, $
H^{0i}_{i}=B_{ij}\psi^{0i}_{j}$ and substituting these into the Eq. (\ref{NReal}), we obtain the explicit Lagrangian responsible for FCNC in the down-sector
\begin{eqnarray}
\mathcal{L}_{FCNC}^{\Delta
S=2}(d)&=&\frac{m_{t}\lambda}{\kappa_{1}\kappa_{2}
-\kappa'_{1}\kappa'_{2}}\bigg(\Big[(\frac{m_{u}}{m_{t}}-\frac{m_{c}}{m_{t}}
)(1-\frac{\lambda^{2}}{2})-A^{2}\lambda^{4}(1-\rho)\Big](\kappa_{2}A^{*}_{i6}
-\kappa'_{2}A^{*}_{i8})H_{i}^{0r}\nonumber\\
&\times&(\bar{d}
P_{R}s+\bar{d}P_{L}s)+A^{2}\lambda^{4}
\eta(\kappa_{2}B^{*}_{i6}-\kappa'_{2}B^{*}_{i8})H_{i}^{0i}(\bar{d}
P_{R}s-\bar{d}P_{L}s)\bigg)\nonumber\\
&+&\frac{i m_{t}\lambda}{\kappa_{1}\kappa_{2}
-\kappa'_{1}\kappa'_{2}}\bigg(\Big[\frac{m_{u}}{m_{t}}-\frac{m_{c}}{m_{t}}
)(1-\frac{\lambda^{2}}{2})-A^{2}\lambda^{4}(1-\rho)\Big](\kappa_{2}B^{*}_{i6}
-\kappa'_{2}B^{*}_{i8})H_{i}^{0i}
\nonumber\\
&\times&(\bar{d}
P_{R}s-\bar{d}P_{L}s)-A^{2}\lambda^{4}
\eta(\kappa_{2}A^{*}_{i6}-\kappa'_{2}A^{*}_{i6})H_{i}^{0r}(\bar{d}
P_{R}s
+\bar{d}P_{L}s)\bigg).
\end{eqnarray}
We proceed in similar fashion to evaluate the flavor-conserving and
flavor-violating Higgs contributions to the up sector. The Yukawa Lagrangian for
the up quark
sector is 
\begin{equation}\label{YukawaUp}
\mathcal{L}^{N}_{Y}(u)=u_{L}^{i*}Y^{ij}_{u}\phi_{1}^{0}u_{R}^{j}+u_{L}^{i*}Y^{ij
}_{d}\chi_{1}^{0}u_{R}^{j}+u_{R}^{j*}\phi_{1}^{0*}Y^{ji*}_{u}u_{L}^{i}+u_{R}^{j*
}\chi_{1}^{0*}Y^{ji*}_{d}u_{L}^{i}.
\end{equation}
We use the same substitutions as before and express the Lagrangian in terms of
the complex  fields $\phi^{0}_{2}$ and $\chi^{0}_{2}$.  
 The first and third terms in the Lagrangian above are flavor-conserving. Writing
the neutral and imaginary parts separately, the FCNC Lagrangian
reads
\begin{eqnarray}\label{NRealUp}
\mathcal{L}_{FCNC}(u)&=&\frac{u_{L}^{n*}V^{nk}\hat{M}^{kk}_{d}
V^{lk*}u_{R}^{l}(\kappa_{2}\phi^{0r}_{1}-\kappa'_{2}\chi^{0r}_{1})}{
\kappa_{1}\kappa_{2}
-\kappa'_{1}\kappa'_{2}}
+\frac{iu_{L}^{n*}V^{nk}\hat{M}^{kk}_{d}V^{lk*}u_{R}^{l}
(\kappa_{2}\phi^{0i}_{1}-\kappa'_{2}\chi^{0i}_{1})}{\kappa_{1}\kappa_{2}
-\kappa'_{1}\kappa'_{2}}\nonumber\\
&+&\frac{u_{R}^{n*}V^{nk}\hat{M}^{kk*}_{d}V^{lk*}u_{R}^{l}
(\kappa_{2}\phi^{0r}_{1}-\kappa'_{2}\chi^{0r}_{1})}{\kappa_{1}\kappa_{2}
-\kappa'_{1}\kappa'_{2}}
-\frac{iu_{R}^{n*}V^{nk}\hat{M}^{kk*}_{d}V^{lk*}u_{R}^{l}
(\kappa_{2}\phi^{0i}_{1}-\kappa'_{2}\chi^{0i}_{1})}{\kappa_{1}\kappa_{2}
-\kappa'_{1}\kappa'_{2}},\nonumber\\
\end{eqnarray}
where $\phi^{0r}_{1}$ and $\chi^{0r}_{1}$ are the two of the nine bare scalar
fields and $\phi^{0i}_{1}$ and $\chi^{0i}_{1}$
are the two of the nine bare pseudo-scalar fields appearing
in LRSUSY Lagrangian. The $u-c$ coupling in Eq. (\ref{NRealUp}) allows
 a $\Delta C=2$ transition at tree level. Inserting 
  $V^{uk}\hat{M}^{kk}_{u}V^{ck*}$  in terms of Wolfenstein parameters,
\begin{equation}
V^{uk}\hat{M}^{kk}_{u}V^{ck*}=(m_{s}-m_{c})(\lambda-\frac{\lambda^{3}}{2})-m_{b}
A^{2}\lambda^{5}(-\rho+i\eta),
\end{equation}
 and using  physical states instead of $\phi^{0r}_{1}$ and
$\chi^{0r}_{1}$ we obtain the explicit form of the Lagrangian responsible for FCNC in the up-sector
\begin{eqnarray}
\mathcal{L}_{FCNC}^{\Delta
C=2}(u)&=&\frac{m_{b}\lambda}{\kappa_{1}\kappa_{2}
-\kappa'_{1}\kappa'_{2}}\bigg(\Big[(\frac{m_{s}}{m_{b}}-\frac{m_{d}}{m_{b}}
)(1-\frac{\lambda^{2}}{2})+A^{2}\lambda^{4}\rho\Big](\kappa_{2}A^{*}_{i5}
-\kappa'_{2}A^{*}_{i7})H^{0r}_{i}\nonumber\\
&\times&(\bar{u}P_{R}c+\bar{u}P_{L}c)+A^{2}\lambda^{4}
\eta(\kappa_{2}B^{*}_{i5}-\kappa'_{2}B^{*}_{i7})H^{0i}_{i}(\bar{u}P_{R}
c-\bar{u}P_{L}c)\bigg)\nonumber\\
&+&\frac{i m_{b}\lambda}{\kappa_{1}\kappa_{2}
-\kappa'_{1}\kappa'_{2}}\bigg(\Big[(\frac{m_{s}}{m_{b}}-\frac{m_{d}}{m_{b}}
)(1-\frac{\lambda^{2}}{2})+A^{2}\lambda^{4}\rho\Big](\kappa_{2}B^{*}_{i5}
-\kappa'_{2}B^{*}_{i7})H^{0i}_{i}\nonumber\\
&\times&(\bar{u}P_{R}c-\bar{u}P_{L}c)-A^{2}\lambda^{4}\eta(\kappa_{2}A^{*}_{i5}
-\kappa'_{2}A^{*}_{i7})H^{0r}_{i}(\bar{u}P_{R}c+\bar{u}P_{L}c)\bigg).
\end{eqnarray}
These expressions will be used to calculate  the real and imaginary parts of the
$K^0- \bar{K}^0,~D^{0}-\bar{D}^{0}$ and $B^{0}-\bar{B}^{0}$ mixing.

\subsection{
\textbf{$\epsilon_K$ and $K^{0}-\bar{K}^{0}$} \textbf{Mixing}}

We evaluate the real and imaginary parts
of the
$K^{0}-\bar{K}^{0}$ transition.  We assume a common mass for scalar and
pseudoscalar Higgs fields.
\begin{eqnarray}
{\rm Re}\langle \bar{K}^{0}|H_{eff}|K^{0}
\rangle&=&\frac{m_{t}^{2}\lambda^{2}}{4M_{i}^{2}(\kappa_{1}\kappa_{2}-\kappa'_{1
}\kappa'_{2})^{2}}\Bigg\{\Big[(\frac{m_{u}}{m_{t}}-\frac{m_{c}}{m_{t}}
)(2-\lambda^ { 2 } )-2A^{2}\lambda^{ 4}(1-\rho)\Big]\nonumber\\
&\times&\bigg([(\kappa_{2}A^{*}_{i6}-\kappa'_{2}A^{*}_{i8})^{2}
-(\kappa_{2}B^{*}_{i6}-\kappa'_{2}B^{*}_{i8})^{2}]
(\langle \tilde{Q}_{1}(\mu)\rangle+\langle
Q_{1}(\mu)\rangle)\nonumber\\
&+&[(\kappa_{2}A^{*}_{i6}-\kappa'_{2}A^{*}_{i8})^{2}
+(\kappa_{2}B^{*}_{i6}-\kappa'_{2}B^{*}_{i8})^{2}]
(\langle \tilde{Q}_{2}(\mu)\rangle+\langle
Q_{2}(\mu)\rangle)\bigg)\nonumber\\
&+&4A^{4}\lambda^{8}\eta^{2}\bigg([(\kappa_{2}A^{*}_{i6}-\kappa'_{2}A^{*}_{i8}
)^{2}+(\kappa_{2}B^{*}_{i6} -\kappa'_{2}B^{*}_{i8})^{2}]
(\langle \tilde{Q}_{1}(\mu)\rangle+\langle
Q_{1}(\mu)\rangle)\nonumber\\
&+&[(\kappa_{2}A^{*}_{i6} -\kappa'_{2}A^{*}_{i8})^{2}-(\kappa_{2}B^{*}_{i6}
-\kappa'_{2}B^{*}_{i8})^{2}]
(\langle \tilde{Q}_{2}(\mu)\rangle+\langle
Q_{2}(\mu)\rangle) \bigg)\Bigg \}, \nonumber\\
\end{eqnarray}
and
\begin{eqnarray}
{\rm Im}\langle
K^{0}|H_{eff}|\bar{K}^{0}\rangle&=&\frac{im_{t}^{2}\lambda^{2}}{4M_{i}^{2}
(\kappa_{1}\kappa_{2}-\kappa'_{1}\kappa'_{2})^{2}}\Big[(\frac{m_{u}}{m_{t}
}-\frac{m_{c}}{m_{t}})(2-\lambda^{2})A^{2}\lambda^{4}
\eta-2A^{4}\lambda^{8}(1-\rho)\eta\Big]\nonumber\\
&\times&\bigg([(\kappa_{2}B^{*}_{i6}-\kappa'_{2}B^{*}_{i8})^{2}-(\kappa_{2}
A^{*}_{i6 }-\kappa'_{2}A^{*}_{i8})^{2}]
(\langle \tilde{Q}_{1}(\mu)\rangle+\langle
Q_{1}(\mu)\rangle)\nonumber\\
&-&[(\kappa_{2}B^{*}_{i6}-\kappa'_{2}B^{*}_{i8})^{2}+(\kappa_{2}
A^{*}_{i6 }-\kappa'_{2}A^{*}_{i8})^{2}]
(\langle \tilde{Q}_{2}(\mu)\rangle+\langle
Q_{2}(\mu)\rangle)\bigg),\nonumber\\
\end{eqnarray}
The quantities $Q_{1}$, $Q_{2}$, $\tilde{Q}_{1}$, and $\tilde{Q}_{2}$
are four quark
operators and are given by
\begin{eqnarray}\label{operators}
Q_{1}&=&(\bar{q}_{1}^{\alpha}P_{L}q_{2}^{\alpha})\otimes(\bar{q}_{1}^{\beta}P_
{ L }q_{2}^{\beta})\qquad,\qquad
\tilde{Q}_{1}=(\bar{q}_{1}^{\alpha}P_{R}q_{2}^{\alpha})\otimes(\bar{q}_{1}^{
\beta}P_{R}q_{2}^{\beta}),\nonumber\\
Q_{2}&=&(\bar{q}_{1}^{\alpha}P_{L}q_{2}^{\alpha})\otimes(\bar{q}_{1}^{\beta}
P_{R} q_{2}^{\beta})\qquad,\qquad
\tilde{Q}_{2}=(\bar{q}_{1}^{\alpha}P_{R}q_{2}^{\alpha})\otimes(\bar{q}
_{1}^{ \beta }P_{L} q_{2}^{\beta}),
\end{eqnarray}
where $\alpha$ and $\beta$ are the color indices. The matrix
elements are, \cite{Buras:2001ra}
\begin{eqnarray}\label{matrixel}
\langle Q_{1}(\mu)\rangle&=&-\frac{5}{24}\bigg(\frac{m_{_{a}}}{m_{q_{1}}
(\mu)+m_{q_{2}}(\mu)}\bigg)^{2}m_{a}F_{a}^{2}B_{1}(\mu),\nonumber\\
\langle Q_{2}(\mu)\rangle&=&\frac{1}{4}\bigg(\frac{m_{a}}{m_{q_{1}}
(\mu)+m_{q_{2}}(\mu)}\bigg)^{2}m_{a}F_{a}^{2}B_{2}(\mu)
\end{eqnarray}
where $a=K, B_{d}, B_{s}, D$ mesons, and  no summation is assumed.  $F_{a}$ is
the decay constant of the corresponding meson and $B_{1}(\mu)$ and $B_{2}(\mu)$
are the
bag parameters calculated in NDR scheme for an energy scale $\mu$. The
numerical values
for all the parameters involved in the calculation of $K^0-{\bar K}^0$,
$D^0-{\bar D}^0$ and
$B_{d,s}^0-{\bar B}_{d,s}^0$ mixings are summarized in Table \ref{tbl:hadronization} and  the quark mass values
in Table III. Same expressions for the
operators $Q_{1}$ and $Q_{2}$ are
valid for the
operators $\tilde{Q}_{1}$ and $\tilde{Q}_{1}$.
\bigskip
\begin{table}[h!]
\centering
\renewcommand{\arraystretch}{1.2}
\small{
\begin{tabular}{|c|c|c|c|c|}
\hline
 & $\mathbf{K^{0}-\bar{K}^{0}}$ & $\mathbf{B_{d}^{0}-\bar{B}_{d}^{0}}$
& $\mathbf{B_{s}^{0}-\bar{B}_{s}^{0}}$ & $\mathbf{D^{0}-\bar{D}^{0}}$\\
\hline\hline
$\mathbf{\mu}$ & $2$ GeV & $m_{b}$ &  $m_{b}$ & $2$ GeV\\
\hline
$\mathbf{q_{1}}$ & $s$ & $b$ & $b$ & $u$\\
\hline
$\mathbf{q_{2}}$ & $d$ & $d$ & $s$ & $c$\\
\hline
$\mathbf{m_{a}}$ & $498$ MeV & $5.28$ GeV & $5.37$ GeV & $1.86$ GeV\\
\hline
$\mathbf{F_{a}}$ & $160$ MeV & $0.21$ GeV & $0.25$ GeV & $232$ MeV\\
\hline
$\mathbf{B_{1}(\mu)}$ & $0.76$ & $0.82$  & $0.83$  & $1$ \\
\hline
$\mathbf{B_{2}(\mu)}$ & $1.30$ & $1.16$  & $1.17$  & $1$ \\
\hline
\end{tabular}}
\caption{QCD parameters used for meson mixing}
\label{tbl:hadronization}
\end{table}
 Substituting  $\mu=2$ GeV in the expressions for $\Delta
M_{K}$ and CP violating parameter $\epsilon_{K}$ given below 
\begin{table}[h!]
\label{tbl:quarks}
\caption{Quark masses}
\begin{center}
\begin{tabular}{c|c|c}
\hline
$m_{u}(2~{\rm GeV})$& $m_{d}(2~{\rm GeV})$ & $m_{s}(2~{\rm GeV})$\\
$2.49^{+0.81}_{-0.79}$ MeV & $5.05^{+0.75}_{-0.95}$ MeV &
$101^{+29}_{-21}$ MeV\\
\hline\hline
$m_{c}(m_{c})$ & $m_{b}(m_{b})$ & $m_{t}(m_{t})$ \\
$1270^{+70}_{-90}$ MeV & $4190^{+180}_{-60}$ MeV & $(172\pm 0.9\pm
1.3) \times 10^3$ MeV\\
\hline
\end{tabular}
\end{center}
\end{table}
\begin{eqnarray}
\Delta
M_{K}=2 {\rm Re}\langle \bar{K}^{0}|H_{eff}|K^{0}
\rangle,~~~~
\Delta
\epsilon_{K}=\frac{1}{\sqrt{2}\Delta M_{K}} {\rm Im}\langle \bar{K}^{0}|H_{eff}|K^{0}
\rangle,
\end{eqnarray}
we get 
\begin{equation}
\Delta
M_{K}=\frac{6.9269\times 10^{-7} A_{i6}^{2*}+2.0088\times
10^{-7}B_{i6}^{2*}}{M_{i}^{2}}(1+ \tan\beta^{2}),
\end{equation}
and 
\begin{eqnarray}\label{epsilonK}
\epsilon_{K}&=&\frac{9.9975\times 10^{6}
A_{i6}^{2*}-9.8616\times 10^{-9}A_{i6}^{*}B_{i6}^{*}
+2.8993\times10^{7}B_{i6}^{2*}}{M_{i}^{2}}(1+ \tan\beta^{2}).
\end{eqnarray}
By comparing the calculated expressions with their experimental values, we
obtain on the sources of flavor and CP violation in the LRSUSY.

The experimental value for the mass difference of $K_{L}$ and $K_{S}$ is
given by \cite{particledata} 
\begin{equation}
|\Delta M_K|=M_{K_{L}}-M_{K_{S}}=(3.483 \pm 0.006)\times 10^{-12} ~{\rm MeV}
\end{equation}
and indirect CP violation in $K\rightarrow \pi\pi $ \cite{Christenson} and in
$K\rightarrow \pi l\nu $ decays is given by \cite{particledata}
\begin{equation}
|\epsilon_{K}|=(2.228 \pm 0.011)\times 10^{-3}
\end{equation}
We  give below the analytical expressions for the constraints on the parameters
in the neutral scalar and pseudoscalar mixing from $K$  meson mixing. Taking
the lightest neutral Higgs mass to be $M_{H^{0r}_{i}}=M_{H^{0i}_{i}}=M_{i}$,
 the value of $\Delta M_{K}=3.483\times 10^{-15}$ GeV yields the constraint
\begin{equation}
M_{i}^{2}\geq(1.9888\times 10^{8} A_{i6}^{2*}+5.7675\times
10^{8}B_{i6}^{2*})(1+ \tan\beta^{2})~{\rm GeV}^{2}
\end{equation}
while the value of $\epsilon_{K}=2.228\times 10^{-3}$ \cite{particledata} yields
the constraint
\begin{equation}
M_{i}^{2}\geq (4.4872\times 10^{9} A_{i6}^{2*}-4.4262\times
10^{-6}A_{i6}^{*}B_{i6}^{*}+1.3013\times
10^{10}B_{i6}^{2*})(1+\tan\beta^{2})~{\rm GeV}^{2}
\end{equation}
In the above expressions we assumed that the lightest Higgs mass provides the dominant contribution, and neglected the rest, while in our numerical evaluations we have summed over all mass contributions, as in  (4.20) and (4.21). These become, for example, when $\tan\beta= 10$
\begin{equation}
M_{i}^{2} \geq (2.0087\times 10^{10} A_{i6}^{2*}+5.8251\times
10^{10}B_{i6}^{2*})~{\rm GeV^{2}}
\end{equation}
and 
\begin{equation}
M_{i}^{2}\geq (4.5320\times 10^{11} A_{i6}^{2*}-4.4704\times
10^{-4}A_{i6}^{*}B_{i6}^{*}+1.3143\times
10^{12}B_{i6}^{2*})~{\rm GeV}^{2}
\end{equation}
%
We  tried varying the lightest relative masses in the scalar and
pseudoscalar sector and found that the results do not change. 
\subsection{\textbf{$B^{0}_{d}-\bar{B}^{0}_{d}$} \textbf{Mixing}}

We proceed the same way as for $K^0-\bar{K}^0$ mixing to evaluate the
constraints from the $B^0_d,~B^0_s$ meson mixing. We use again four quark
operators $Q_{1}$, $Q_{2}$, $\tilde{Q}_{1}$, and $\tilde{Q}_{2}$ defined previously. Setting
as before the Higgs mass to be equal to the lightest scalar mass $M_{H^{0r}_{i}}=M_{H^{0i}_{i}}=M_{i}$
 the expression for $\Delta
M_{B_{d}}$ becomes
\begin{equation}
\Delta M_{B_{d}}=\frac{(9.4139\times
10^{-6}A_{i6}^{2*}+3.6405\times
10^{-5}B_{i6}^{2*})(1+\tan^{2}\beta)}{M^{2}_{i}}~{\rm GeV}^{3}
\end{equation}
Using the experimental value of $\Delta M_{B_{d}}=3.337\times 10^{-13}~{\rm
GeV}$ \cite{particledata}, we obtain, assuming as before dominance by the lightest mass
\begin{equation}
M_{i}^{2}\geq (2.8211\times 10^{7} A_{i6}^{2*}+1.6909\times
10^{8}B_{i6}^{2*})(1+\tan^{2}\beta)~{\rm GeV}^{2}
\end{equation}
which becomes, for $\tan\beta= 10$
\begin{equation}
M_{i}^{2}\geq (2.8493\times 10^{9}A_{i6}^{2*}+1.1019\times
10^{10}B_{i6}^{2*})~{\rm GeV}^{2}
\end{equation}

\subsection{\textbf{$B^{0}_{s}-\bar{B}^{0}_{s}$} \textbf{Mixing}}

We proceed exactly as in the previous subsection, substituting $s$ instead
of $d$ quark. The parameters for $B_{s}^0-\bar{B}_{s}^0$ mixing  are given
in Table \ref{tbl:hadronization}.
\begin{equation}
\Delta M_{B_{s}}=\frac{(4.2314\times 10^{-4}
A_{i6}^{*^{2}}+1.6469\times
10^{-3}B_{i6}^{*^{2}})(1+\tan^{2}\beta)}{M_{i}^{2}}~{\rm GeV}^{3}
\end{equation}
Using the experimental value of $\Delta M_{B_{d}}=117\times 10^{-13}~{\rm GeV}$
\cite{particledata, CDFD0}
\begin{equation}
M_{i}^{2}\geq (3.6166\times 10^{7} A_{i6}^{2*}+1.4076\times
10^{8}B_{i6}^{2*})(1+\tan^{2}\beta)~{\rm GeV}^{2}
\end{equation}
or, for $\tan\beta= 10$
\begin{equation}
M_{i}^{2}\geq (3.6528\times 10^{9} A_{i6}^{2*}+1.4217\times
10^{10}B_{i6}^{2*})~{\rm GeV}^{2}
\end{equation}
%
\subsection{\textbf{$D^{0}-\bar{D}^{0}$} \textbf{Mixing}}

In  subsection {\bf A}, we evaluated the real and imaginary parts of the
$D^{0}-\bar{D}^{0}$ transition. We assume as before a common mass for scalar and
pseudo-scalar Higgs fields.
\begin{eqnarray}
{\rm Re}\langle \bar{D}^{0}|H_{eff}|D^{0}
\rangle&=&\frac{m_{b}^{2}\lambda^{2}}{4M_{i}^{2}
(\kappa_{1}\kappa_{2}-\kappa'_{1}\kappa'_{2})^{2}}\Bigg\{\Big[(\frac{m_{s}}
{m_{b}}-\frac{m_{d}}{m_{b}})(2-\lambda^{2})+2A^{2}\lambda^{4}\rho\Big]^{2}
\nonumber\\
&\times&\bigg([(\kappa_{2}A^{*}_{i5}-\kappa'_{2}A^{*}_{i7})^{2}-(\kappa_{2}B^{*}
_ {i5} -\kappa'_ {2}B^{*}_ { i7 })^{2}]
(\langle \tilde{Q}_{1}(\mu)\rangle+\langle
Q_{1}(\mu)\rangle)\nonumber\\
&+&[(\kappa_{2}A^{*}_{i5}-\kappa'_{2}A^{*}_{i7})^{2}+(\kappa_{2}B^{*}
_ {i5} -\kappa'_ {2}B^{*}_ { i7 })^{2}]
(\langle \tilde{Q}_{2}(\mu)\rangle+\langle
Q_{2}(\mu)\rangle)\bigg)\nonumber\\
&+&4A^{4}\lambda^{8}\eta^{2}\bigg([(\kappa_{2}A^{*}_{i5}-\kappa'_{2}A^{*
}_{i7})^{2}+(\kappa_{2}B^{*}
_ {i5} -\kappa'_ {2}B^{*}_ { i7 })^{2}]
(\langle \tilde{Q}_{1}(\mu)\rangle+\langle
Q_{1}(\mu)\rangle)\nonumber\\
&+&[(\kappa_{2}A^{*}_{i5}-\kappa'_{2}A^{*
}_{i7})^{2}-(\kappa_{2}B^{*}
_ {i5} -\kappa'_ {2}B^{*}_ { i7 })^{2}]
(\langle \tilde{Q}_{2}(\mu)\rangle+\langle
Q_{2}(\mu)\rangle\bigg)\Bigg\}\nonumber\\
\end{eqnarray}
where $Q_{1}$, $Q_{2}$, $\tilde{Q}_{1}$, and $\tilde{Q}_{2}$
are the four quark
operators defined as before,
the mass difference $\Delta
M_{D}=2 {\rm Re}\langle \bar{D}^{0}|H_{eff}|D^{0}
\rangle$ is obtained as
\begin{equation}
\Delta M_{D}=\frac{5.2816\times 10^{-10} A_{i7}^{2*}+5.8097\times
10^{-9}B_{i7}^{2*}}{M_{i}^{2}}\frac{(1+\tan\beta^{2})}{\tan\beta^{2}}~{\rm
GeV}^{3}.
\end{equation}
Comparing the calculated expression with the experimental
value  \cite{particledata}
\begin{equation}
|\Delta M_D|=M_{D_{1}^{0}}-M_{D_{2}^{0}}=(1.57313)\times 10^{-17}
~{\rm MeV},
\end{equation}
we obtain
\begin{eqnarray}
M_{i}^{2}&\geq&\frac {(3.3574\times 10^{10} A_{i7}^{2*}+3.6931\times
10^{11}B_{i7}^{2*})(1+ \tan\beta^{2})}{\tan\beta^{2}}~{\rm
GeV}^{2},
\end{eqnarray}
which becomes for $\tan \beta=10$,
\begin{eqnarray}
M_{i}^{2}&\geq&(3.3909\times 10^{10} A_{i7}^{2*}+3.7300\times
10^{11}B_{i7}^{2*})~{\rm GeV}^{2}.
\end{eqnarray}
\section{Numerical Results and Discussion}
\label{discussion}
The FCNCs tree-level diagrams are mediated by the physical scalar
fields $H_{3}^{0}$ and $H_{9}^{0}$, and the pseudoscalars
$A_{2}^{0}$ and $A_{7}^{0}$. These fields are linear superpositions of  the
$\chi_1^{0r}$ or $\phi_2^{0r}$ ($\chi_1^{0i}$ or $\phi_2^{0i}$ , respectively,
for the pseudoscalars) components from the bidoublet Higgs.  

As the fields $H_{3}^{0}$ and $H_{9}^{0}$ must be heavy, the light neutral
scalars would likely be linear combinations of the complimentary $\chi_2^{0r}$ or
$\phi_1^{0r}$ components from the bidoublets, or $\delta^{0r},
\bar{\delta}^{0r}, \delta^{c^{0r}}$, and $\bar{\delta}^{c^{0r}}$ from the triplet
Higgs.
 We set $v_R$ in the interval obtained from the requirement that the doubly
charged Higgs are light ($3-10$ TeV). Varying $v_R$ outside this range adversely
affects the masses of the lightest
doubly charged Higgs, and some of the light neutral and singly charged scalars.
 
 The mass of the lightest scalar field $H_{1}^{0}$ (the SM-like) changes at most
a
few GeV, if we vary any of the parameters, whereas the second
lightest scalar field $H_{2}^{0}$ is highly dependent on the changes in the
parameter $v_{R}$. Both of these fields can be light, as our numerical explorations indicate. Similarly, the lightest pseudoscalar field $A_{1}^{0}$
behaves like the second lightest neutral scalar field and 
is also effected by the changes in $v_{R}$. $H_1^0$ is SM-like, and the
parameter that seems to affect $H_{1}^{0}$ mass the most is the $\lambda_{21}$
coupling. (This parameter is the coupling that generates the $\mu_{21}=\lambda_{21} \langle S \rangle$ Higgsino coupling). The dependence is not smooth, but varying $\lambda_{21}$ in the
interval $0.01-1$ produces a 30\% change in $M_{H_{1}^{0}}$.

\begin{figure}[htb]
\hspace*{-1.1cm}
	\includegraphics[width=2.2in,height=2.2in]{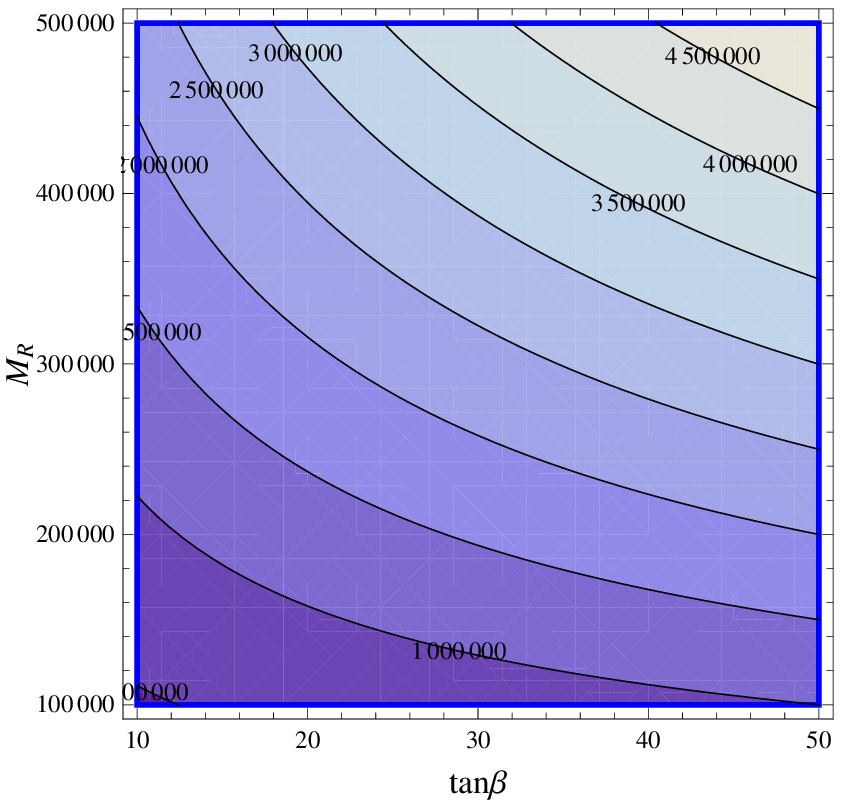}
	\includegraphics[width=2.2in,height=2.2in]{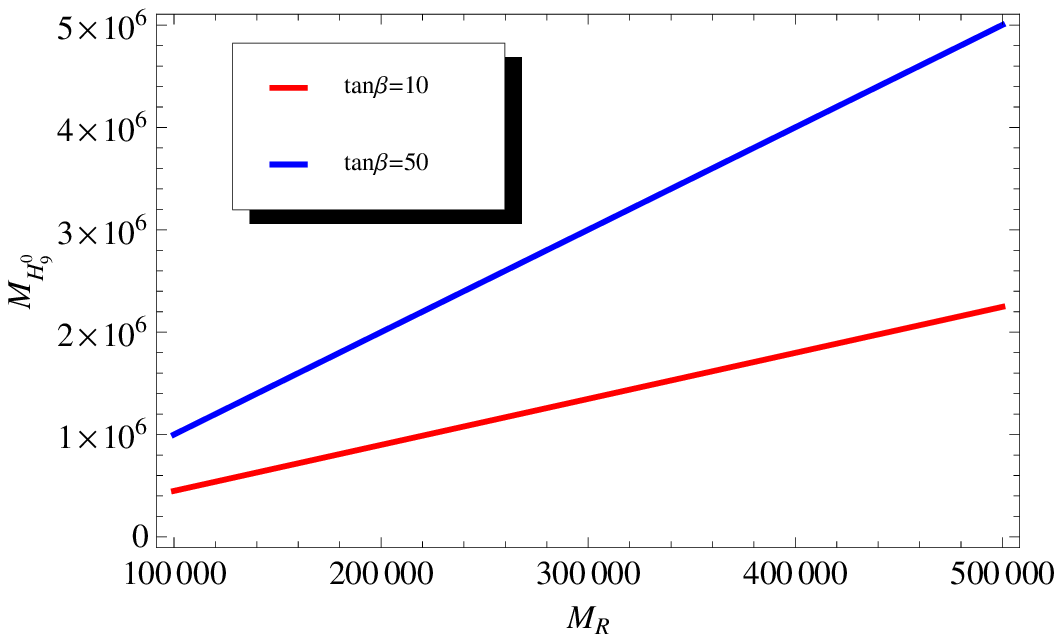}
\hspace*{-0.2cm}
	\includegraphics[width=2.2in,height=2.2in]{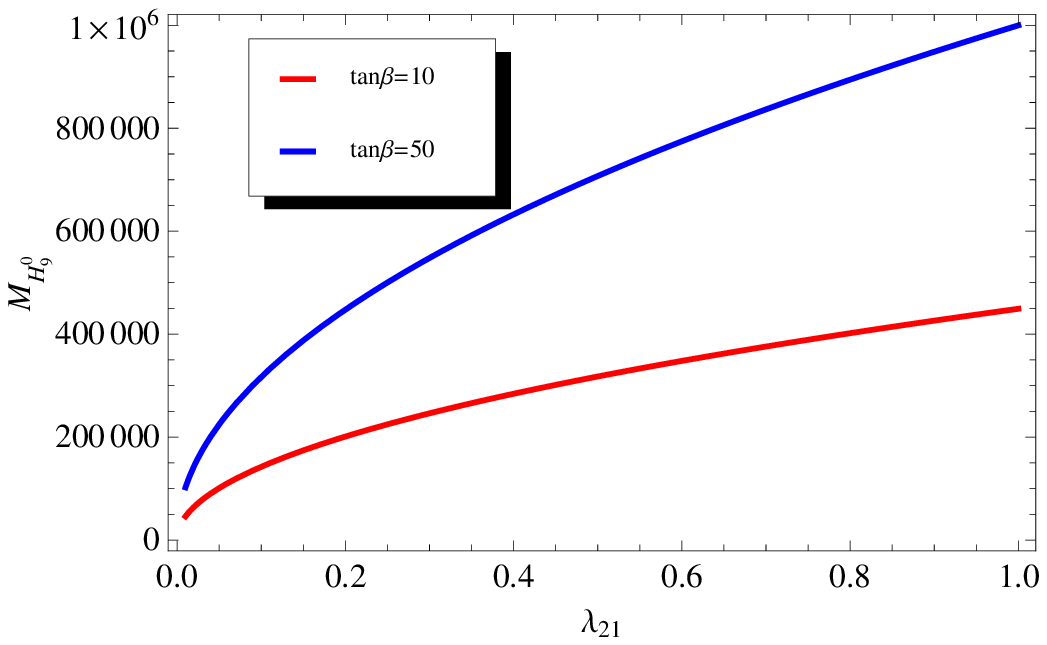}
\vskip -0.3in
      \caption{The variation of the FCNC neutral Higgs $H^0_9$ mass with the
parameters of the LRSUSY model. $H^0_9$ induces tree-level FCNC in the
down-quark sector. Shown are: contour plots in the $M_R-\tan \beta$ plane, 
the variation of $M_{H^0_9}$ with $M_R$, and with $\lambda_{21}$,  for
two values of $\tan \beta$. Masses are given in GeV.}
\label{fig:H9Higgs}
\end{figure}

The tree-level flavor-changing neutral currents in the down-quark sector are
governed by $H_{9}^{0}$ and $A_{7}^{0}$. The mass values of the fields
$H_{9}^{0}$ and $A_{7}^{0}$ are the same, and they are 
dependent on the parameters $\lambda_{21}~, v_R,~\lambda,~\tan \beta $ and
$M_R$. Numerical investigation reveals that only $\tan \beta,~M_R$ and
$\lambda_{21}$ can affect the $H_{9}^{0}$ and $A_{7}^{0}$ masses significantly. For instance,  if
$\lambda_{21}$ increases from $0.01$ to $1$, the $H_9^0$ mass 
increases almost $10$ times. %
These masses are also slightly dependent
on the parameters $M_{R}$ and $\tan\beta$ such that when they
increase, mass values of these physical fields also increase. The
dependence of the $H_{9}^{0}$ mass on the parameter $M_{R}$ is more dominant
than on $\tan\beta$. Requiring $M_R \sim 100$ TeV insures that Higgs-mediated FCNCs in K and B neutral mesons are suppressed to levels consistent with experimental data. The variations of $H^0_9$ mass with these parameters are shown in Fig.~\ref{fig:H9Higgs}. 

\begin{figure}[htb]
\begin{center}$
	\begin{array}{cc}
\hspace*{-1.1cm}
	\includegraphics[width=3.0in,height=3.0in]{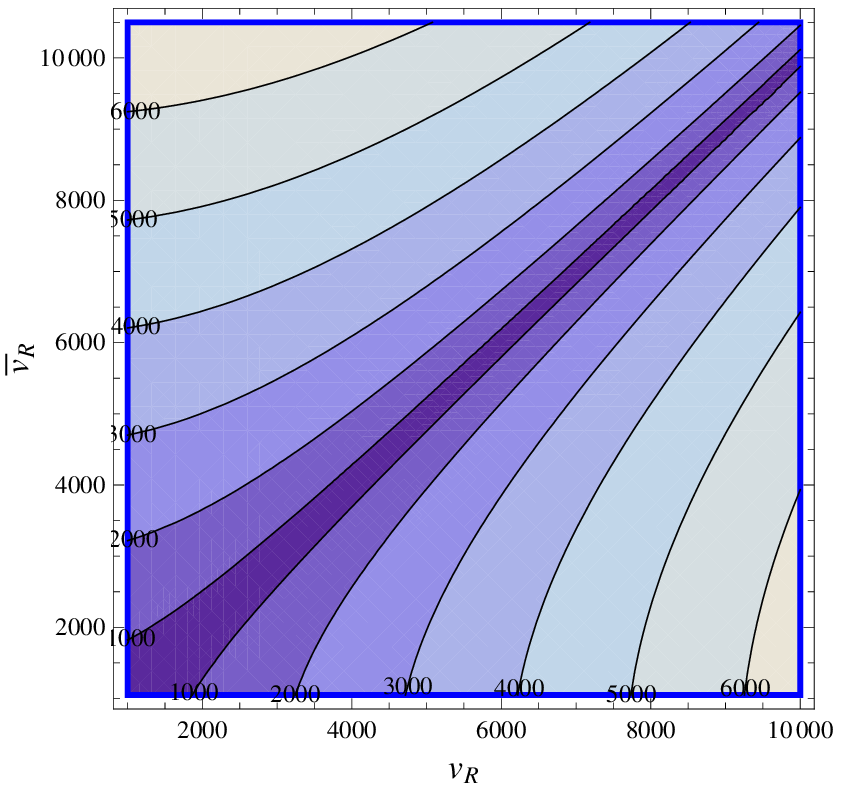}
&\hspace*{-0.2cm}
	\includegraphics[width=3.0in,height=3.0in]{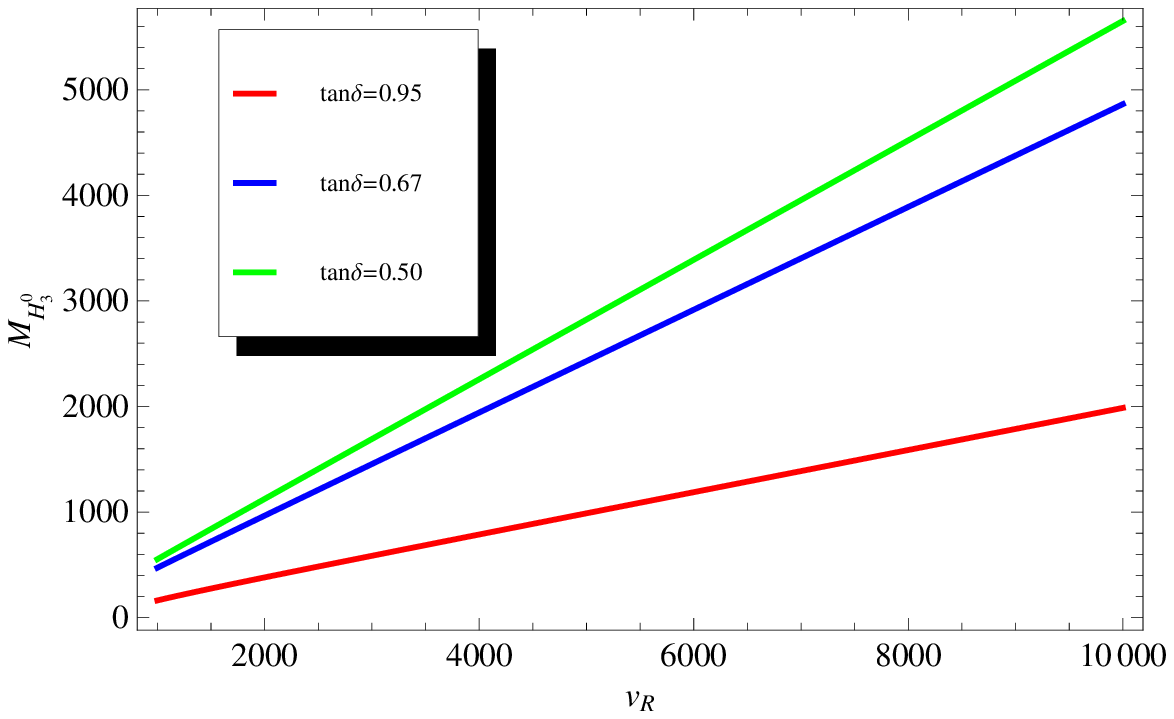} 
\end{array}$
\end{center}
\vskip-0.3in
 \caption{The variation of the FCNC neutral Higgs $H^0_3$ mass with the
parameters of the LRSUSY model. $H^0_3$ induces tree-level FCNC in the up-quark
sector. To the left,  a contour plot in the $v_R-\bar{v}_R$ plane and, at the right, as a function of $v_R$ for three values of $\tan \delta={\bar{v}_R}/{v_R}$. Masses are given in GeV.}
 \label{fig:H3Higgs}
\end{figure}

The  fields
$H_{3}^{0}$ and $A_{2}^{0}$ are responsible for flavor-changing neutral currents
in the up-quark sector. Their masses are the same (as one can infer from the
mass matrices in Section {\ref{results}}), and although they depend in principle
on $v_R,~\tan \beta$ and $\lambda_{21}$, the only significant dependence is on
$v_R$, such that if $v_{R}$ increases from $3$
 to $10$ TeV, their mass values increase approximately $5$
times. The mass also varies with the ratio $\tan \delta={\bar{v}_R}/{v_R}$,  while almost independent of the changes in the other
parameters. The parameter dependence is shown in Fig.~\ref{fig:H3Higgs}, where we plot the explicit $v_R$ dependence for three values of $\tan \delta$, as well as a more extensive illustration of the $v_R-\bar{v}_R$ dependence in a contour plot.   $D^0-\bar{D}^0$ mixing constraints require $v_R \ge 3$ TeV. While the dependence on both $\tan \beta$ and $\lambda_{21}$ is very weak,  the dependence on $v_R$ is almost linear. 
%

\begin{figure}[htb]
\hspace*{-1.1cm}
	\includegraphics[width=3.0in,height=3.0in]{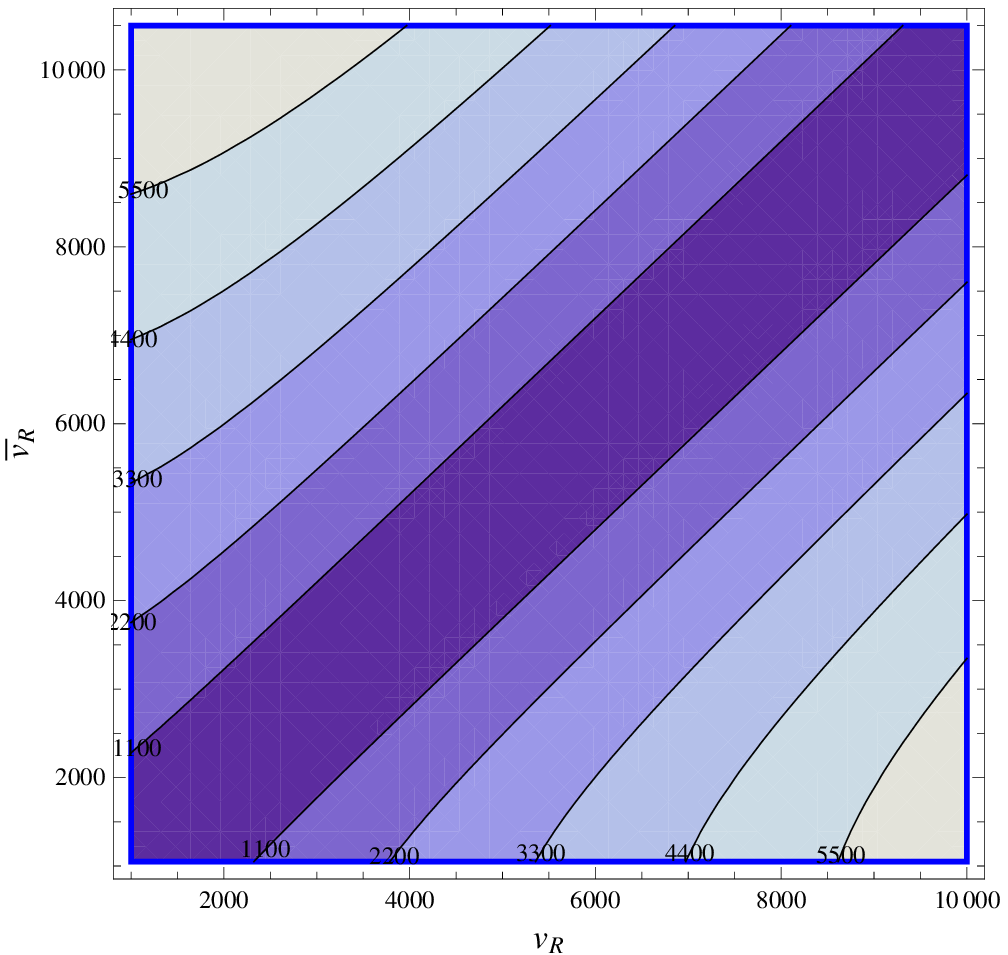}
	\includegraphics[width=3.0in,height=3.0in]{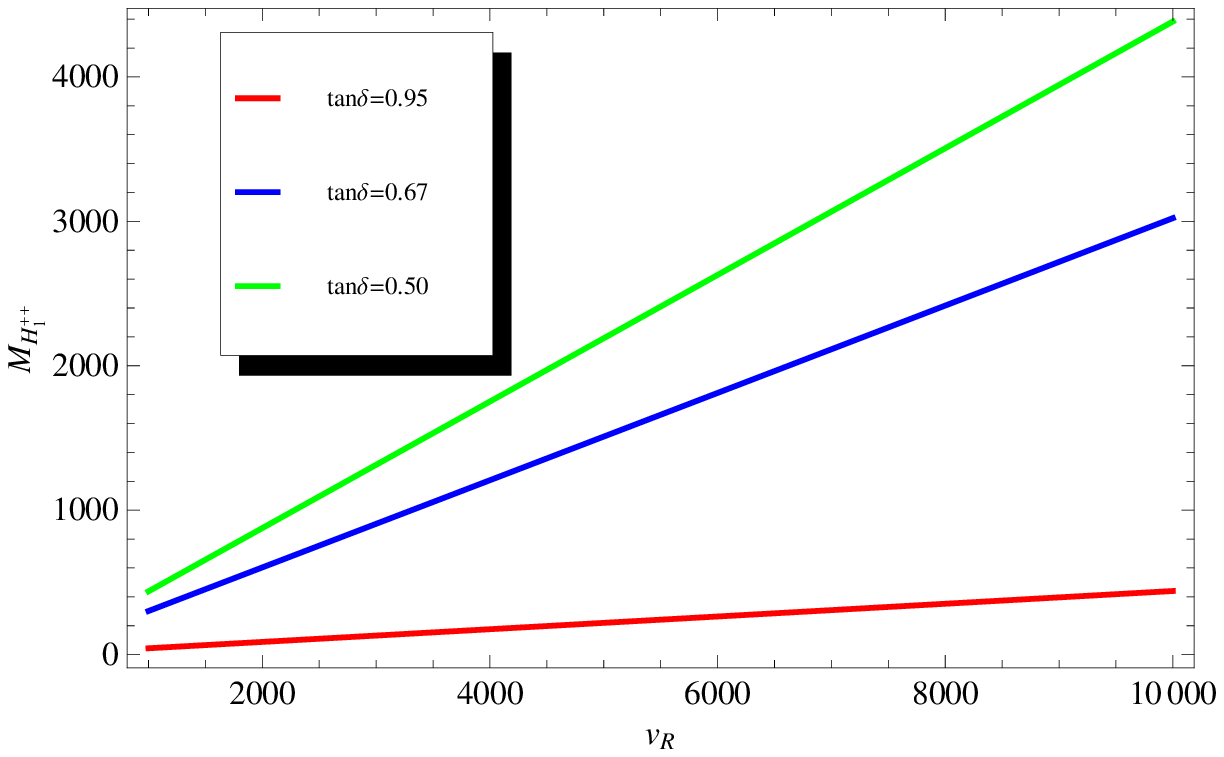} 
\vskip-0.3in
 \caption{The masses of the lightest doubly-charged Higgs boson  as a contour plot in the $v_R-\bar{v}_R$ plane (left) and as a function of $v_R$ for three values of $\tan \delta={\bar{v}_R}/{v_R}$ (right). Masses are given in GeV.}
\label{fig:doublyHiggs}
\end{figure}

From the approximate analytical expressions in Section ({\ref{results}}), the
mass of the lightest doubly charged physical field
$H_{1}^{\pm \pm}$ depends on $v_R, ~\lambda$ and $M_R$. Analysis shows that only  the
dependence on $v_R$ is significant. However, the exact mass also depends on $\bar{v}_R$ through the ratio $\tan \delta={\bar{v}_R}/{v_R}$. As before we show, in Fig.~\ref{fig:doublyHiggs}, the dependence of these two parameters as a contour plot in the $v_R-\bar{v}_R$ plane. The mass of $H_1^{\pm \pm}$ increases with
the increasing values of $v_{R}$ , as shown on the right hand side of Fig.~\ref{fig:doublyHiggs},  for three values of $\tan \delta$, while it is basically independent on $M_R$.  One can see that for $v_R \sim 3.5 $ TeV the doubly charged Higgs boson mass is light for all values of $\tan \delta$, while for  $v_R=10$ TeV the mass is highly dependent on ${\bar{v}_R}/{v_R}$.
For example, when we change $v_{R}$ from $3$ to $10$ TeV,
the $H_1^{\pm \pm}$ mass values increase approximately $4$ times. The effect of
varying the other parameters is negligible for the lightest doubly charged
Higgs, whereas the mass of the heavier doubly charged Higgs $H_2^{\pm \pm}$
depends almost exclusively on $M_R$.

The lightest singly charged physical field
$H_{1}^{\pm}$ mass can also depend in principle on $v_R,~\lambda,~\tan \beta$
and $M_R$. However, upon inspection, the only significant dependence is on
$v_R$, much like the doubly charged Higgs boson. The reason is that, requiring one doubly charged Higgs to be light makes the lightest singly charged Higgs to be an eigenstate of the $2 \times 2$ mass matrix $M^{2}_{\delta^{+}\bar{\delta}^{-*}}$.   There is a slight difference between the lightest singly charged and lightest doubly charged Higgs boson, but the difference comes from SM-like observables in $g_L^2\kappa_{dif}^2$, and it is overwhelmed by parameters proportional to $v_R$ and $M_R$. Its mass  also increases with
the increasing values of $v_{R}$, and depends on the ratio $\bar{v}_R/v_R$ in much the same way as the mass of the  doubly charged Higgs does. Thus we do not show the dependence separately.

Finally, we present two explicit numerical scenarios for the Higgs masses, which
obey the constraints from meson mixings: one for $v_R=3.5$ TeV and $\tan \beta=10$, the other for
$v_R=5$ TeV and $\tan \beta=50$. The other parameters in both scenarios are taken to be $\tan\delta \equiv {\bar{v}_R}/{v_R}=1/1.05$,
$M_{R}=100$ TeV, $\lambda=1$, $\lambda_{21}=1$,
$C_{\lambda}=2.5$ TeV, $\langle S\rangle=1$ TeV, $M_{S}=1$
TeV. We give masses and compositions in terms of the bare states. One can see that, except for raising the lightest neutral Higgs mass, increasing $\tan \beta$ has little effect on the spectrum. However raising $v_R$ increases the mass of the lighter non-SM-like Higgs bosons in the neutral scalar and pseudoscalar sector, as well as in the singly and doubly charged Higgs sectors.
While we did not prove in general that the model conserves $R$-parity, the
numerical results  obtained from minimizing the masses confirm the results of
\cite{Babu:2008ep}. Both of these scenarios allow for a pair of light flavor-conserving neutral scalar Higgs bosons (one SM-like, one mostly triplet $SU(2)_L$); as well as for one light singly charged Higgs and a pair of doubly charged
Higgs bosons. The FCNC Higgs responsible for mixing in the up ($D^0-\bar{D}^0$)
or down ($K^0-\bar{K}^0$ and $B_{d,s}^0-\bar{B}_{d,s}^0$) quark sectors are
heavy and satisfy the experimental constraints in each sector. 

\setlength{\voffset}{-0.5in}
\begin{table}[h!]
\centering
\small{
\begin{tabular}{|l|r|l|}
\hline
Particle & Mass (GeV) & Composition\\
\hline\hline
$H_{1}^{0}$ & $100.6$ &
$0.015\delta^{c^{0r}}+0.014\bar{\delta}^{c^{0r}}+0.099\phi_{1}^{0r}+0.995\chi^{
0r}_ {2}-0.010S^{0r}$ \\
$H_{2}^{0}$ & $151.9$ & $0.724\delta^{0r}+0.690\bar{\delta}^{0r}$ \\
$H_{3}^{0}$ & $680.9$ & $0.100\phi^{0r}_{2}+1.000\chi_{1}^{0r}$\\
$H_{4}^{0}$ & $3433.4$ &
$-0.522\delta^{c^{0r}}-0.497\bar{\delta}^{c^{0r}}+0.002\phi^{0r}_{1}+0.032\chi^{
0r}_{2}+0.693S^{0r}
$\\
$H_{5}^{0}$ & $5997.1$ &
$0.502\delta^{c^{0r}}+0.478\bar{\delta}^{c^{0r}}-0.001\phi^{0r}_{1}-0.007\chi^{
0r}_{2}
+0.721 S^{0r}$\\
$H_{6}^{0}$ & $141419.5$ &
$-0.690\delta^{0r}+0.724\bar{\delta}^{0r}$\\
$H_{7}^{0}$ & $141537.9$ &
$-0.690\delta^{c^{0r}}+0.724\bar{\delta}^{c^{0r}}$\\
$H_{8}^{0}$ & $449294.1$ & $-1.000\phi^{0r}_{1}+0.100\chi_{2}^{0r}$\\
$H_{9}^{0}$ & $449294.8$ &  $1.000\phi^{0r}_{2}-0.100\chi^{0r}_{1}$ \\
\hline
$A_{1}^{0}$ & $151.9$ & $0.724\delta^{0i}-0.690\bar{\delta}^{0i}$ \\
$A_{2}^{0}$ & $680.9$ & $-0.100\phi^{0i}_{2}+1.000\chi_{1}^{0i}$ \\
$A_{3}^{0}$ & $4935.7$ & $1.000S^{0i}$\\
$A_{4}^{0}$ & $141419.5$ & $-0.690\delta^{0i}-0.724\bar{\delta}^{0i}$\\
$A_{5}^{0}$ & $141502.0$ & $0.690\delta^{c^{0i}}+0.724\bar{\delta}^{c^{0i}}$ \\
$A_{6}^{0}$ & $449294.1$ & $1.000\phi^{0i}_{1}+0.100\chi_{2}^{0i}$ \\
$A_{7}^{0}$ & $449294.8$ & $1.000\phi^{0i}_{2}+0.100\chi^{0i}_{1}$\\
\hline
$H_{1}^{+}$ & $152.9$ & $0.724\delta^{+}+0.690\bar{\delta}^{-*}$ \\
$H_{2}^{+}$ & $690.2$ &
$-0.018\delta^{c^{-*}}-0.018\bar{\delta}^{c^{+}}-0.099\phi_{2}^{-*}
+0.995\chi_ {1}^{+}$ \\
$H_{3}^{+}$ & $141419.5$ &
$-0.690\delta^{+}+0.724\bar{\delta}^{-*}$
\\
$H_{4}^{+}$ & $141454.7$ &
$0.690\delta^{c^{-*}}-0.724\bar{\delta}^{c^{+}}$\\
$H_{5}^{+}$ & $449294.3$ & $0.995\phi_{1}^{+}+0.100\chi_{2}^{-*}$ \\
$H_{6}^{+}$ &  $449294.8$ & $0.995\phi^{-*}_{2}+0.100\chi_{1}^{+}$ \\
\hline
$H_{1}^{++}$ & $153.9$ &
$0.724\delta^{++}+0.690\bar{\delta}^{--*}$\\
$H_{2}^{++}$ & $216.3$
& $0.724\delta^{c^{--*}}+0.690\bar{\delta}^{c^{++}}$ \\
$H_{3}^{++}$ & $141419.5$ &
$-0.690\delta^{++}+0.724\bar{\delta}^{--*}$ \\
$H_{4}^{++}$ & $141419.6$ &
$-0.690\delta^{c^{--*}}+0.724\bar{\delta}^{c^{++}}$ \\
\hline
$G_{1}^{0}$ & $0$ &
$-0.721\delta^{c^{0i}}+0.686\bar{\delta}^{c^{0i}}+0.010\phi^{0i}_{1}
-0.095\chi_{2}^{0i}$\\
$G_{2}^{0}$ & $0$
&$0.069\delta^{c^{0i}}-0.066\bar{\delta}^{c^{0i}}+0.099\phi_{1}^{0i}
-0.990\chi_{2}^{0i}$ \\
$G_{1}^{+}$ & $0$ & $0.100\phi_{1}^{+}-0.995\chi_{2}^{-*}$\\
$G_{2}^{+}$ & $0$
&$0.724\delta^{c^{-*}}+0.690\bar{\delta}^{c^{+}}-0.003\phi_{2}^{-*}
+0.025\chi_ { 1 }^{+}$\\
\hline
\end{tabular}}
\caption{Masses and compositions of physical
Higgs fields and unphysical Goldstone bosons. Parameters are chosen
as follows: $\tan\beta=10$, $\tan\delta \equiv {\bar{v}_R}/{v_R}=1/1.05$,
$v_{R}=3.5$ TeV,
$M_{R}=100$ TeV, $\lambda=1$, $\lambda_{21}=1$,
$C_{\lambda}=2.5$ TeV, $\langle S\rangle=1$ TeV, $M_{S}=1$
TeV.}
\label{tbl:quarkmasstanten} 
\end{table}

\setlength{\voffset}{-0.5in}
\begin{table}[h!]
\centering
\small{
\begin{tabular}{|l|r|l|}
\hline
Particle & Mass (GeV) & Composition\\
\hline\hline
$H_{1}^{0}$ & $112.9$ &
$0.002\delta^{c^{0r}}+0.002\bar{\delta}^{c^{0r}}+0.020\phi_{1}^{0r}+1.000\chi^{
0r}_ {2}-0.001S^{0r}$ \\
$H_{2}^{0}$ & $218.5$ & $0.724\delta^{0r}+0.690\bar{\delta}^{0r}$ \\
$H_{3}^{0}$ & $998.6$ & $0.020\phi^{0r}_{2}+1.000\chi_{1}^{0r}$\\
$H_{4}^{0}$ & $5562.6$ &
$-0.522\delta^{c^{0r}}-0.497\bar{\delta}^{c^{0r}}+0.005\chi^{0r}_{2}+0.693S^{0r}
$\\
$H_{5}^{0}$ & $8901.0$ &
$0.519\delta^{c^{0r}}+0.494\bar{\delta}^{c^{0r}}-0.003\chi^{0r}_{2}+0.697
S^{0r}$
\\
$H_{6}^{0}$ & $141333.6$ &
$-0.690\delta^{0r}+0.724\bar{\delta}^{0r}$\\
$H_{7}^{0}$ & $141575.2$ &
$-0.690\delta^{c^{0r}}+0.724\bar{\delta}^{c^{0r}}$\\
$H_{8}^{0}$ & $999258.8$ & $-1.000\phi^{0r}_{1}+0.020\chi^{0r}_{2}$ \\
$H_{9}^{0}$ & $999258.3$ & $1.000\phi^{0r}_{2}-0.020\chi_{1}^{0r}$ \\
\hline
$A_{1}^{0}$ & $218.5$ & $0.724\delta^{0i}-0.690\bar{\delta}^{0i}$ \\
$A_{2}^{0}$ & $998.6$ & $-0.020\phi^{0i}_{2}+1.000\chi_{1}^{0i}$ \\
$A_{3}^{0}$ & $6976.8$ & $1.000S^{0i}$\\
$A_{4}^{0}$ & $141334.6$ & $-0.690\delta^{0i}-0.724\bar{\delta}^{0i}$\\
$A_{5}^{0}$ & $141502.0$ & $-0.690\delta^{c^{0i}}-0.724\bar{\delta}^{c^{0i}}$ \\
$A_{6}^{0}$ & $999252.8$ & $1.000\phi^{0i}_{1}+0.020\chi^{0i}_{2}$\\
$A_{7}^{0}$ & $999258.3$ & $1.000\phi^{0i}_{2}+0.020\chi_{1}^{0i}$ \\
\hline
$H_{1}^{+}$ & $219.2$ & $0.724\delta^{+}+0.690\bar{\delta}^{-*}$ \\
$H_{2}^{+}$ & $995.3$ &
$0.013\delta^{c^{-*}}+0.012\bar{\delta}^{c^{+}}+0.020\phi_{2}^{-*}
-1.000\chi_ {1} ^{+}$ \\
$H_{3}^{+}$ & $141334.6$ &
$-0.690\delta^{+}+0.724\bar{\delta}^{-*}$
\\
$H_{4}^{+}$ & $141405.3$ &
$0.690\delta^{c^{-*}}-0.724\bar{\delta}^{c^{+}}$\\
$H_{5}^{+}$ &  $999258.3$ & $1.000\phi^{+}_{1}+0.020\chi_{2}^{-*}$ \\
$H_{6}^{+}$ & $999259.8$ & $1.000\phi_{2}^{-*}+0.020\chi_{1}^{+}$ \\
\hline
$H_{1}^{++}$ & $219.9$ &
$0.724\delta^{++}+0.690\bar{\delta}^{--*}$\\
$H_{2}^{++}$ & $310.2$ &
$0.724\delta^{c^{--*}}+0.690\bar{\delta}^{c^{++}}$ \\
$H_{3}^{++}$ & $141334.6$ &
$-0.690\delta^{c^{--*}}+0.724\bar{\delta}^{c^{++}}$ \\
$H_{4}^{++}$ & $141334.7$ &
$-0.690\delta^{++}+0.724\bar{\delta}^{--*}$ \\
\hline
$G_{1}^{0}$ & $0$ &
$-0.200\delta^{c^{0i}}+0.190\bar{\delta}^{c^{0i}}-0.019\phi^{0i}_{1}
+0.961\chi_{2}^{0i}$\\
$G_{2}^{0}$ & $0$
&$0.696\delta^{c^{0i}}-0.663\bar{\delta}^{c^{0i}}-0.006\phi_{1}^{0i}+0.276\chi_
{2}^{0i}$ \\
$G_{1}^{+}$ & $0$ & $0.020\phi_{1}^{+}-1.000\chi_{2}^{-*}$\\
$G_{2}^{+}$ & $0$
&$0.724\delta^{c^{-*}}+0.690\bar{\delta}^{c^{+}}-0.001\phi_{2}^{-*}
+0.018\chi_ { 1 }^{+}$\\
\hline
\end{tabular}}
\caption{Masses and compositions of
physical Higgs fields and unphysical Goldstone bosons. Parameters
are chosen as follows: $\tan\beta=50$, $\tan\delta=1/1.05$,
$v_{R}=5$ TeV, $M_{R}=100$ TeV, $\lambda=1$,
$\lambda_{21}=1$, $C_{\lambda}=2.5$ TeV, $\langle S\rangle=1$
TeV, $M_{S}=1000$ GeV.}
\label{tbl:quarkmasstanfifty}
\end{table}

 \section{Summary and Conclusion}
 \label{conclusion}

We analyzed the Higgs sector of a minimal left-right supersymmetric model with
automatic $R$-parity violation. Symmetries of the model forbid explicit
$R$-parity violation. Inclusion of the effects of the Yukawa coupling of the
heavy Majorana neutrino insures a global minimum which is charge conserving,
thus avoiding spontaneous $R$-parity breaking or the need to introduce higher
dimensional terms.

The Higgs sector contains four doubly charged Higgs, six singly charged Higgs
fields,  nine neutral scalar fields, and seven pseudoscalar fields (in addition
to two neutral Goldstone bosons, and two charged ones).  One would expect that,
with so many free  parameters in the Lagrangian, and so many free masses, almost
any scenario is possible for the Higgs masses in this model. We show that the
requirement that 1) there is a light neutral scalar Higgs boson, flavor
conserving, which is the counterpart to the SM Higgs boson; 2) there exist at
least one light doubly charged Higgs boson (as it is interesting for
phenomenology); and 3) the flavor-violating neutral Higgs bosons satisfy the
constraints imposed by the experimental data from $K^0-\bar{K}^0$, 
$D^0-\bar{D}^0$, and $B_{d,s}^0-\bar{B}_{d,s}^0$ mixings, makes the Higgs sector fairly
predictive and fixes some of the parameters in a narrow range. The masses of the
light neutral and doubly charged Higgs bosons depend on very few parameters. For
instance, we find that requirement 1) and 2) are related, and satisfied by $v_R
\in (3, 10)$ TeV range. Assuming $v_R \sim \bar{v}_R$ and $g_L=g_R$, this
predicts masses for the $W_R$ around $4 - 13$ TeV (assuming negligible mixing
with $W_L$), and for $Z_R$ bosons in the  $3 - 10$ TeV range. Thus, while the model can allow for light neutral, singly and doubly charged Higgs bosons, it predicts new gauge bosons  just outside the range $M_{W_R} <2 (4)$ TeV which can be observed at the LHC with a luminosity of 0.1(30) fb$^{-1}$ \cite{Ferrari:2000sp}.

The parameter $M_R$, associated with the singlet Higgs field in the
superpotential, must be of $\cal{O}$(100) TeV, which insures high masses for 
the FCNC Higgs.

Our analysis is important for two reasons: first, we have shown that a
reasonable Higgs mass spectrum is possible in LRSUSY, without all Higgs masses
being required to be heavy. We can require that the Higgs generating tree-level
FCNC in the $K,~D$ and $B$ mesons are heavy, but still obtain two light 
neutral Higgs bosons, one light pseudoscalar,  one light singly charged Higgs boson, and a pair of light doubly charged Higgs bosons. Second,  as most
Higgs masses are sensitive to few parameters, the model is very predictive and free of
additional parameters, such as the sneutrino VEVs or extra higher-dimensional terms. Third, the non-SM light Higgs are mostly triplet $SU(2)_L$ bosons and expected to decay copiously to leptons, but not to quarks, giving clear distinguishing signals for the model. 
This analysis can now form the basis of a consistent phenomenological study of
signals from such a Higgs sector, including production and decay rates, and has
implications for the masses of the additional gauge bosons, as well as for the
right-handed neutrinos.  
\section{Acknowledgments}
 We thank NSERC of Canada for partial financial support under 
Grant No. SAP01105354 and Alper Hayreter for discussions. 


%

\end{document}